\documentclass[aps,prd,twocolumn,nofootinbib,superscriptaddress,longbibliography]{revtex4-2}

\usepackage[T1]{fontenc}
\usepackage[utf8]{inputenc}
\usepackage{lmodern}
\usepackage{amsmath,amssymb,mathtools,bm}
\usepackage{physics}
\usepackage{hyperref}
\usepackage{microtype}
\usepackage{xcolor}
\usepackage{enumitem}
\usepackage{tensor}

\hypersetup{colorlinks=true,linkcolor=blue,citecolor=blue,urlcolor=blue}

\newcommand{\TT}{\mathrm{TT}}
\newcommand{\TF}{\mathrm{TF}}
\newcommand{\Dij}{D_{ij}}
\newcommand{\phib}{\bar\phi}
\newcommand{\rhob}{\bar\rho_\phi}
\newcommand{\Pb}{\bar P_\phi}
\newcommand{\omegab}{\omega(\bar\phi)}
\newcommand{\omegapb}{\omega_{,\phi}(\bar\phi)}
\newcommand{\omegappb}{\omega_{,\phi\phi}(\bar\phi)}
\newcommand{\Vb}{V(\bar\phi)}
\newcommand{\Vpb}{V_{,\phi}(\bar\phi)}
\newcommand{\Vppb}{V_{,\phi\phi}(\bar\phi)}
\newcommand{\Lie}{\pounds}
\newcommand{\Tmath}{\mathfrak T}

\begin{document}

\title{Thermal channels of scalar and tensor waves in Jordan-frame scalar--tensor gravity}

\author{David S. Pereira}
\email{djpereira@fc.ul.pt}

\author{Francisco S.N Lobo}%
	\email{fslobo@fc.ul.pt}
\author{Jos\'e Pedro Mimoso}
\email{jpmimoso@fc.ul.pt}

	\affiliation{%
		Departamento de F\'{i}sica, Faculdade de Ci\^{e}ncias da Universidade de Lisboa, Campo Grande, Edif\'{\i}cio C8, P-1749-016 Lisbon, Portugal}
	\affiliation{Instituto de Astrof\'{\i}sica e Ci\^{e}ncias do Espa\c{c}o, Faculdade de
		Ci\^encias da Universidade de Lisboa, Campo Grande, Edif\'{\i}cio C8,
		P-1749-016 Lisbon, Portugal;\\
	}%
\date{\today}

\begin{abstract}
We study first-order scalar and tensor perturbations of Jordan-frame scalar--tensor gravity about a spatially flat FLRW background using the Einstein-like effective-fluid decomposition of the scalar sector. In the scalar-gradient frame, we derive the perturbed effective density, pressure, heat flux, and anisotropic stress, and show that they admit an exact Eckart-type constitutive identification at linear order. We then show that these same quantities appear explicitly and exhaustively in the linearized field equations: the scalar Hamiltonian, momentum, trace, and traceless Einstein-like equations are governed, respectively, by the effective density, heat-flux, pressure, and anisotropic-stress channels, while the tensor propagation equation is governed by the transverse-traceless anisotropic-stress channel. In particular, the Jordan-frame modification of gravitational-wave damping is identified with the effective transverse-traceless anisotropic stress of the scalar sector. We also derive the perturbed evolution equation for the invariant product $\kappa T$, clarify its gauge behavior, and show that flux matching on FLRW fixes only the background value $\overline{\kappa T}$, not its perturbation. These results leave open the possibility that gravitational waves in scalar--tensor gravity admit a deeper thermodynamic characterization, perhaps even an intrinsic one, although the present analysis establishes this only at the level of an effective constitutive description.
\end{abstract}

\maketitle


\section{Introduction}

Scalar--tensor gravity is one of the best-motivated and most extensively studied extensions of general relativity~\cite{Brans:1961sx,Bergmann:1968ve,Wagoner:1970vr,Fujii:2003pa,Kase:2019veo,Kobayashi:2019hrl,Ishak:2018his,Nojiri:2017ncd,Kase:2018aps,CANTATA:2021asi,CosmoVerseNetwork:2025alb}. In these theories the gravitational interaction is mediated not only by the metric, but also by an additional scalar degree of freedom, which modifies the relation between geometry and matter and opens a wide phenomenological arena ranging from solar-system tests and compact objects to inflation, dark energy, and large-scale cosmology~\cite{Faraoni:2004pi}. In the Jordan-frame formulation, the scalar field enters directly in the gravitational sector through a nonminimal coupling to curvature, so that the effective gravitational coupling becomes dynamical. This formulation is particularly useful because the modified field equations can be rewritten in an Einstein-like form, with the non-Einstein contributions grouped into an effective source built out of the scalar field and its derivatives.

That rewriting has long been exploited as an organizational tool in gravitation and cosmology, but in recent years it has acquired a more specific interpretation: the effective scalar sector can be decomposed as an imperfect fluid carrying an energy density, isotropic pressure, heat flux, and anisotropic stress \cite{Faraoni:2021lfc,Faraoni:2025alq,Faraoni:2025dex,Faraoni:2025fjq,Gallerani:2025myd,Karolinski:2024nwp,Giardino:2023qlu,Faraoni:2023hwu,Giardino:2022sdv,Faraoni:2021jri,Karolinski:2024ukr,Faraoni:2022gry,Bhattacharyya:2025tgp,Miranda:2024dhw,Giusti:2021sku,Houle:2024sxs,Giusti:2026ymb,Miranda:2022wkz,Faraoni:2022doe}. In this formulation, the scalar sector exhibits a nontrivial transport-like structure and naturally invites a thermodynamic or constitutive interpretation. In particular, one can identify effective heat-flux and anisotropic-stress channels, discuss their relation to Eckart-type constitutive laws, and define a natural gravity-side thermal variable through the invariant combination $\kappa T$ with $\kappa$ being the effective conductivity and $T$ effective temperature of the effective scalar sector. Even if one remains agnostic about any microscopic thermodynamic interpretation, this effective-fluid viewpoint already captures genuine structural information about the scalar--tensor dynamics.

At the same time, the perturbative sector of scalar--tensor gravity contains genuine propagating degrees of freedom: the usual transverse-traceless tensor modes and, in general, an additional scalar mode~\cite{Hwang:1996xh,Hwang:1995bv,Carloni:2006gy,Carloni:2006fs,Kase:2018aps}. Their cosmological propagation equations are well known and have been studied extensively in both the classical scalar--tensor literature and in more general Horndeski-type settings~\cite{Hwang:1996xh,Kobayashi:2019hrl,Kase:2018aps}. Yet these perturbative equations are usually discussed independently of the effective-fluid decomposition. What remains much less explored is whether the full linearized perturbation problem can itself be organized directly in terms of effective thermal channels, and, if so, which combinations of effective density, pressure, heat flux, and anisotropic stress actually control scalar and tensor propagation.

The aim of this paper is to address these questions in Jordan-frame scalar--tensor gravity on a spatially flat FLRW background. Our approach is deliberately computation driven: rather than postulating a thermodynamic analogy and applying it heuristically, we derive the relevant perturbed geometric and field-theoretic quantities explicitly and only thereafter identify the constitutive structure encoded in the resulting equations. This strategy allows us to separate statements that follow directly from linear perturbation theory from broader interpretive claims. In particular, we do not assume that the scalar--tensor sector is microscopically thermal; instead, we ask to what extent the perturbation equations themselves admit a precise organization in terms of thermal channels.

More generally, the link between anisotropic stress and modified gravitational-wave propagation is known in broad classes of modified-gravity theories, including scalar--tensor models~\cite{Saltas:2014dha}. The result established here is more specific. For Jordan-frame scalar--tensor gravity about a spatially flat FLRW background, we show that this relation admits an exact first-order realization in effective thermodynamic variables. In particular, the modified tensor damping term is identified with the transverse-traceless anisotropic-stress channel of the effective scalar sector, while the scalar perturbation system can be reconstructed channel by channel in terms of an effective energy density, pressure, heat flux, and anisotropic stress.

Our main result is that, on flat FLRW, the perturbative effective density, pressure, heat flux, and anisotropic stress of Jordan-frame scalar--tensor gravity admit an exact Eckart-type constitutive identification in the scalar-gradient frame, and that this constitutive structure is realized explicitly in the full first-order perturbation system. More precisely, the scalar Hamiltonian, momentum, trace, and traceless Einstein-like equations are governed, respectively, by the effective density, heat-flux, pressure, and anisotropic-stress channels of the scalar sector, while the standard Jordan-frame gravitational-wave equation is governed by the transverse-traceless anisotropic-stress channel of that same sector. 

To our knowledge, this precise perturbative thermodynamic matching of the full linearized system has not been exhibited previously in Jordan-frame scalar--tensor gravity. The novelty does not lie in the mere appearance of density, pressure, momentum/heat-flux, and anisotropic-stress slots in a projected Einstein-like system, since that is a generic consequence of the $1+3$ decomposition. Rather, the nontrivial statement is that these channels possess a precise Eckart-type constitutive meaning and appear explicitly and exhaustively in the complete first-order scalar and tensor field equations. In this sense, the effective-fluid description is not merely a suggestive rewriting of the background equations, but an exact organizational structure for the perturbative dynamics.

We work with the Einstein-like form of the Jordan-frame field equations and project the effective scalar contribution with respect to a timelike congruence. Whenever it is well defined, we adopt the future-directed scalar-gradient frame, which is the natural frame for the effective thermodynamic interpretation developed in recent work \cite{Faraoni:2021lfc,Faraoni:2025alq,Faraoni:2025dex,Faraoni:2025fjq,Gallerani:2025myd,Houle:2024sxs,Karolinski:2024nwp,Giardino:2023qlu,Faraoni:2023hwu}. Because the relevant quantities involve projections, gradients, frame-dependent decompositions, and gauge-sensitive perturbations, a careful perturbative treatment is essential. We therefore work within the Stewart--Walker--Bruni--Nakamura framework \cite{Stewart:1974uz,Bruni:1996im,Sonego:1997np,Nakamura:2004rm,Malik:2008im} and specialize to scalar-plus-tensor perturbations in Newtonian gauge on flat FLRW. This setting is sufficiently simple to allow an explicit derivation of all effective thermal variables, while still retaining the full scalar and tensor propagation content.

The paper is organized as follows. Section~\ref{sec:JFSTT} reviews the Jordan-frame field equations and introduces the Einstein-like effective source. Section~\ref{sec:backgroundPertSetup} presents the FLRW background, the perturbation framework, and the scalar-plus-tensor ansatz. In Sec.~\ref{sec:thermalVars}, we derive the perturbed effective thermal variables. Section~\ref{sec:EckartConnection} develops the Eckart-type constitutive interpretation, including the projected-temperature-gradient analysis and the entropy discussion. In Sec.~\ref{sec:deltaKTevolution}, we derive the perturbed heating/cooling law for $\kappa T$, analyze its gauge properties, and explain why flux matching on FLRW does not determine $\delta(\kappa T)$. Section~\ref{sec:EinsteinSideChannels} rewrites the perturbed Einstein-like equations in terms of thermal channels, with Subsec.~\ref{sec:scalarpart} devoted to the scalar part of the metric field equations and Subsec.~\ref{sec:tensorwave} to tensor gravitational waves and the transverse-traceless thermal channel. In Sec.~\ref{sec:scalarwave} we study the perturbed scalar-wave equation and its connection with the effective thermal variables. Finally, Sec.~\ref{sec:Discussion} summarizes the physical interpretation and discusses possible extensions.

\section{Jordan-frame scalar--tensor gravity and effective-fluid decomposition}\label{sec:JFSTT}

We consider the Jordan-frame action
\begin{eqnarray}
	S&=&\frac{1}{16\pi}\int d^4x\sqrt{-g}\left[\phi R-\frac{\omega(\phi)}{\phi}\nabla_a\phi\nabla^a\phi-2V(\phi)\right] \nonumber\\ 
	&&\qquad +S_m[g_{ab},\Psi_m].
	\label{eq:action}
\end{eqnarray}
which describes a scalar--tensor theory in which the scalar field $\phi$ is nonminimally coupled to the Ricci scalar $R$, and $\omega(\phi)$ is a (possibly field-dependent) coupling function. The term $V(\phi)$ represents a self-interaction potential for the scalar field, while $S_m$ encodes the matter sector, minimally coupled to the metric $g_{ab}$.

Variation with respect to the metric gives
\begin{eqnarray}
	G_{ab}
	&=&\frac{8\pi}{\phi}T^{(m)}_{ab}
	+\frac{\omega(\phi)}{\phi^2}\left(\nabla_a\phi\nabla_b\phi-\frac12 g_{ab}(\nabla\phi)^2\right)\nonumber\\
	&&\quad +\frac{1}{\phi}\left(\nabla_a\nabla_b\phi-g_{ab}\Box\phi\right)-\frac{V(\phi)}{\phi}g_{ab},
	\label{eq:metricFE}
\end{eqnarray}
which can be interpreted as generalized Einstein equations with an effective gravitational coupling $1/\phi$. Besides the standard matter contribution $T^{(m)}_{ab}$, the scalar field contributes through kinetic, potential, and second-derivative terms, reflecting its dynamical role in the gravitational interaction.

Variation with respect to the scalar field yields the scalar-field equation
\begin{equation}
	(2\omega+3)\Box\phi = 8\pi T^{(m)}-\omega_{,\phi}(\nabla\phi)^2+2\phi V_{,\phi}-4V.
	\label{eq:scalarFE}
\end{equation}
This equation governs the dynamics of $\phi$, showing that it is sourced by the trace of the matter energy--momentum tensor $T^{(m)}$, as well as by its own self-interactions through $\omega(\phi)$ and $V(\phi)$.

It is convenient to define an Einstein-like effective source through
\begin{equation}\label{eq:EinsteinLikeTmathfrak}
	G_{ab}=8\pi\,\Tmath_{ab},
	\qquad
	\Tmath_{ab}\equiv \frac{1}{\phi}T^{(m)}_{ab}+T^{(\phi)}_{ab},
\end{equation}
so that the field equations take a form formally identical to Einstein's equations, but with a modified total energy--momentum tensor $\Tmath_{ab}$.

The scalar-field contribution is explicitly given by
\begin{eqnarray}
	8\pi T^{(\phi)}_{ab}
	&=&\frac{\omega(\phi)}{\phi^2}\left(\nabla_a\phi\nabla_b\phi-\frac12 g_{ab}(\nabla\phi)^2\right)\nonumber\\
	&& +\frac{1}{\phi}\left(\nabla_a\nabla_b\phi-g_{ab}\Box\phi\right)-\frac{V}{\phi}g_{ab}.
	\label{eq:TphiDef}
\end{eqnarray}
This effective tensor collects all contributions from the scalar degree of freedom, including kinetic, potential, and derivative coupling terms, allowing one to interpret deviations from general relativity as arising from an effective fluid associated with $\phi$.

Given a timelike unit vector $u^a$, the effective source is decomposed as
\begin{equation}
	T^{(\phi)}_{ab}=\rho_\phi u_a u_b + P_\phi h_{ab}+2u_{(a}q^{(\phi)}_{b)}+\pi^{(\phi)}_{ab},
	\label{eq:fluidDecomp}
\end{equation}
which corresponds to the standard $1+3$ covariant decomposition of an energy--momentum tensor into effective fluid variables. Here $\rho_\phi$ and $P_\phi$ represent the energy density and isotropic pressure associated with the scalar field, while $q_a^{(\phi)}$ and $\pi^{(\phi)}_{ab}$ encode energy flux and anisotropic stresses, respectively.

These quantities are obtained through the projections
\begin{align}
	\rho_\phi &= T^{(\phi)}_{ab}u^a u^b, \label{eq:rhoProj}\\
	P_\phi &= \frac13 h^{ab}T^{(\phi)}_{ab}, \label{eq:PProj}\\
	q_a^{(\phi)} &= - h_a{}^c T^{(\phi)}_{cd}u^d, \label{eq:qProj}\\
	\pi^{(\phi)}_{ab} &= \left(h_a{}^c h_b{}^d-\frac13 h_{ab}h^{cd}\right)T^{(\phi)}_{cd}, \label{eq:piProj}
\end{align}
where each expression isolates a distinct physical component of the effective fluid by projecting along and orthogonally to $u^a$.

The projection tensor onto the instantaneous rest space orthogonal to $u^a$ is defined as
\begin{equation}
	h_{ab}=g_{ab}+u_a u_b,
\end{equation}
and satisfies $h_{ab}u^b=0$. The vector $u^a$ is interpreted as the 4-velocity of the effective fluid associated with the scalar field.

We will consider the future-directed scalar-gradient frame wherever $X\equiv -(\nabla\phi)^2>0$, namely
\begin{equation}
	u_a=\frac{\nabla_a\phi}{\sqrt{X}},
	\label{eq:gradientFrame}
\end{equation}
which aligns the fluid flow with the gradient of the scalar field. In this frame, the scalar field behaves as a comoving clock. On the FLRW background considered below, this choice implies $\bar u^a=(1,0,0,0)$ and $\bar u_a=(-1,0,0,0)$, consistently with a homogeneous and isotropic cosmological setting.

Additionally, we introduce the 4-acceleration
\begin{equation}
	a_a \equiv u^b\nabla_b u_a,
	\label{eq:accDef}
\end{equation}
which measures the failure of $u^a$ to follow geodesics, and the full velocity-gradient tensor together with the associated expansion scalar
\begin{equation}
	\Theta_{ab} = h_a{}^c h_b{}^d\nabla_{(c}u_{d)},\quad \Theta=\nabla_c u^c\,,
\end{equation}
where parentheses denote symmetrization, $\nabla_{(c}u_{d)} \equiv \tfrac12(\nabla_c u_d+\nabla_d u_c)$. The tensor $\Theta_{ab}$ is the expansion tensor while $\Theta$ represents the volume expansion rate of the effective fluid congruence.

\section{Perturbations, gauge maps, background and observer choice}\label{sec:backgroundPertSetup}

\subsection{Perturbations and gauge maps}

For the perturbations, we will use the standard framework for relativistic perturbation theory \cite{Stewart:1974uz,Bruni:1996im,Sonego:1997np,Nakamura:2004rm,Malik:2008im}. Let $\{(\mathcal M_\epsilon,g_{ab}(\epsilon),\phi(\epsilon),\Psi_m(\epsilon))\}$ be a one-parameter family of spacetimes and fields
with background $(\mathcal M_0,\bar g_{ab},\bar\phi,\bar\Psi_m)$ at $\epsilon=0$.
This construction allows one to treat perturbations in a fully covariant manner, separating physical effects from coordinate artifacts.

A \emph{gauge choice} is an identification map (diffeomorphism)
$\mathcal X_\epsilon:\mathcal M_0\to\mathcal M_\epsilon$ with $\mathcal X_0=\mathrm{id}_{\mathcal M_0}$.
All perturbations are defined on $\mathcal M_0$ by pulling back:
for any tensor field $Q(\epsilon)$,
\begin{equation}
	\bar Q \equiv Q(0),
	\qquad
	\delta Q \equiv \left.\frac{d}{d\epsilon}\,\mathcal X_\epsilon^\ast Q(\epsilon)\right|_{\epsilon=0}.
	\label{eq:deltaDefPullback}
\end{equation}
In this way, all perturbed quantities are compared at the same spacetime point of the background manifold, making the perturbative expansion well-defined.

Changing the identification map $\mathcal X_\epsilon\to\mathcal Y_\epsilon$ induces a diffeomorphism
$\Phi_\epsilon\equiv \mathcal X_\epsilon^{-1}\circ\mathcal Y_\epsilon:\mathcal M_0\to\mathcal M_0$
generated at first order by a vector field $\xi^a$ on $\mathcal M_0$, and yields the first-order gauge rule
\begin{equation}
	\delta Q \to \delta Q + \Lie_\xi \bar Q,
	\label{eq:gaugeRuleGeneral}
\end{equation}
where $\Lie_\xi$ is the Lie derivative \cite{Stewart:1974uz,Bruni:1996im,Nakamura:2004rm}. This expression shows explicitly how perturbations change under infinitesimal coordinate transformations.

In particular,
\begin{align}
	\delta g_{ab} &\to \delta g_{ab}+\Lie_\xi \bar g_{ab}
	=\delta g_{ab}+2\bar\nabla_{(a}\xi_{b)},
	\label{eq:gaugeMetric}\\
	\varphi\equiv \delta\phi &\to \varphi+\Lie_\xi\bar\phi
	=\varphi+\xi^a\bar\nabla_a\bar\phi,
	\label{eq:gaugeScalar}
\end{align}
so that both metric and scalar perturbations acquire gauge-dependent contributions determined by $\xi^a$.

On an FLRW background, where $\bar\phi=\bar\phi(t)$ depends only on cosmic time, the scalar-field transformation simplifies to $\varphi\to \varphi+\xi^0\dot{\bar\phi}$. This makes explicit that time reparametrizations (encoded in $\xi^0$) directly shift the scalar perturbation, a key feature in constructing gauge-invariant variables.

Once a gauge $\mathcal X_\epsilon$ is fixed, we suppress the pullback $\mathcal X_\epsilon^\ast$ and regard all fields as defined
on the background manifold $\mathcal M_0$. This allows us to work directly with perturbed quantities without explicitly tracking the map between manifolds, with the understanding that all objects have been consistently pulled back.

We work systematically to first order in the perturbative parameter, neglecting all $O(\epsilon^2)$ contributions, and adopt the standard shorthand
\begin{equation}
	g_{ab}(\epsilon)=\bar g_{ab}+\delta g_{ab},
	\qquad
	\phi(\epsilon)=\bar\phi+\varphi,
	\label{eq:epsFamilyDef}
\end{equation}
where $\delta g_{ab}=\left.\frac{d}{d\epsilon}g_{ab}(\epsilon)\right|_{\epsilon=0}$ and
$\varphi=\left.\frac{d}{d\epsilon}\phi(\epsilon)\right|_{\epsilon=0}$ are defined in the pulled-back sense introduced above. In this expansion, barred quantities denote background fields, while unbarred perturbations represent first-order deviations.

For the background spacetime, we specialize to a spatially flat Friedmann--Lemaître--Robertson--Walker (FLRW) geometry,
\begin{equation}
	ds^2=-dt^2+a^2(t)\delta_{ij}dx^i dx^j,
	\qquad
	\phi(t,\vb{x})=\phib(t),
	\label{eq:FLRWbg}
\end{equation}
which is homogeneous and isotropic at zeroth order. Due to Eq.~\eqref{eq:gradientFrame} and the future-directed choice one has $\dot{\bar{\phi}}<0$. Here $a(t)$ is the scale factor, and the scalar field depends only on cosmic time at the background level.

We also introduce the Hubble parameter $H\equiv \dot a/a$, which characterizes the expansion rate of the universe and will play a central role in the dynamics of both the background and perturbations.

\subsection{Metric and perturbation variables}

We adopt the scalar-plus-tensor Newtonian gauge ($B=E=0$) metric
\begin{equation}
	ds^2=-(1+2A)dt^2+a^2\left[(1-2\psi)\delta_{ij}+\chi_{ij}\right]dx^i dx^j,
	\label{eq:metricPert}
\end{equation}
which provides a convenient and physically transparent parametrization of perturbations. In this gauge, the scalar degrees of freedom are entirely captured by the potentials $A$ and $\psi$, which coincide (up to sign conventions) with the gauge-invariant Bardeen potentials \cite{Bardeen:1980kt,Kodama:1984ziu,Ma:1994dv,Malik:2008im}. The tensor perturbation $\chi_{ij}$ describes gravitational waves and satisfies the transverse-traceless conditions
\begin{equation}
	\partial^i\chi_{ij}=0,
	\qquad
	\chi^i{}_i=0,
\end{equation}
ensuring that it carries only the two physical tensor polarizations. At linear order on an FLRW background, this tensor sector is automatically gauge invariant \cite{Bardeen:1980kt,Kodama:1984ziu,Malik:2008im}.

For the scalar field, the perturbation is introduced as
\begin{equation}
	\phi(t,\vb{x})=\phib(t)+\varphi(t,\vb{x}),
	\label{eq:phipert}
\end{equation}
where $\varphi$ represents the first-order fluctuation around the homogeneous background value. 

The scalar-gradient frame~\eqref{eq:gradientFrame} yields 
\begin{equation}
	\delta u_0 = -A,
\end{equation}
showing that the temporal component of the velocity perturbation is directly tied to the lapse perturbation. For the spatial components, one finds
\begin{equation}
	\delta u_i=-\frac{\partial_i\varphi}{\dot\phib}\equiv \partial_i v_\phi,
	\qquad
	v_\phi\equiv -\frac{\varphi}{\dot\phib},
	\label{eq:duiBranch}
\end{equation}
assuming $\dot\phib\neq 0$ and selecting the future-directed branch, which implies $\dot{\phi}<0$. This relation identifies $v_\phi$ as the scalar velocity potential associated with the effective fluid description of the scalar field.

For the effective fluid $4$-acceleration we have the perturbed expression (see~\ref{app:perturbed31} for details)
\begin{equation}
	\delta a_i=
	\partial_i\left(A+\dot v_\phi\right),
	\label{eq:deltaaiStart}
\end{equation}
which shows that the acceleration is sourced by both the gravitational potential and the time variation of the velocity potential.

The perturbed expansion scalar, $\delta \Theta$, is given by
\begin{equation}
	{
		\delta\Theta
		=
		\frac{\nabla^2}{a^2}\,v_\phi
		-3\dot\psi
		-3H A.
	}
	\label{eq:deltaTheta}
\end{equation}
This quantity measures the perturbation in the local volume expansion rate and receives contributions from velocity gradients as well as from the scalar metric perturbations.

These variables will be useful both in the scalar momentum channel and in the Eckart constitutive rewriting, where the effective fluid interpretation of the scalar field dynamics becomes particularly transparent.

\section{Perturbed imperfect fluid quantities}\label{sec:thermalVars}

The perturbed effective variables are obtained by perturbing Eqs.~\eqref{eq:rhoProj}--\eqref{eq:qProj} and computing explicitly the contribution of each term. This procedure follows directly from the $1+3$ covariant decomposition introduced earlier, applied at first order in perturbation theory. The intermediate steps are relegated to Appendix~\ref{app:thermalDerivations}; here we collect the final results in forms suitable for use in the Einstein-like perturbation equations.

Starting with the energy density $\rho_{\phi}$ and the pressure $P_\phi$, their first-order perturbations are obtained from the corresponding projections of $T^{(\phi)}_{ab}$. One finds
\begin{equation}
	\delta\rho_\phi = \delta T^{(\phi)}_{00}-2\rhob A,
	\label{eq:deltaRhoProjection}
\end{equation}
and
\begin{equation}
	\delta P_\phi = \frac{1}{3a^2}\delta^{ij}\delta T^{(\phi)}_{ij}+2\Pb\psi,
	\label{eq:deltaPProjection}
\end{equation}
where the additional terms proportional to $A$ and $\psi$ arise from the perturbation of the projection tensors. The background quantities entering these expressions are given by
\begin{align}
	8\pi\rhob&=\frac{\omegab}{2}\left(\frac{\dot\phib}{\phib}\right)^2-3H\frac{\dot\phib}{\phib}+\frac{\Vb}{\phib},
	\label{eq:rhobg}\\
	8\pi\Pb&=\frac{\omegab}{2}\left(\frac{\dot\phib}{\phib}\right)^2+\frac{\ddot\phib+2H\dot\phib}{\phib}-\frac{\Vb}{\phib}.
	\label{eq:Pbg}
\end{align}
These expressions correspond to the effective background energy density and pressure associated with the scalar field.

Carrying out the explicit computation of the perturbations yields
\begin{eqnarray}
	8\pi\,\delta\rho_\phi
	&=&
	\frac{\omegab}{\phib^2}\dot\phib\dot\varphi
	-\frac{\omegab}{\phib^2}\dot\phib^{2}A \nonumber\\ 
	&&+\frac{1}{\phib}\left(\frac{\nabla^2}{a^2}\varphi-3H\dot\varphi+3\dot\phib\dot\psi+6H\dot\phib A\right)\nonumber\\
	&&+\left(\frac{\omegapb}{2\phib^2}-\frac{\omegab}{\phib^3}\right)\dot\phib^{2}\varphi
	+\frac{3H\dot\phib}{\phib^2}\varphi \nonumber \\
	&&+\left(\frac{\Vpb}{\phib}-\frac{\Vb}{\phib^2}\right)\varphi,
	\label{eq:deltaRhoFinal}
\end{eqnarray}
and
\begin{eqnarray}
	8\pi\,\delta P_\phi
	&=&
	\frac{\ddot\varphi}{\phib}
	+\frac{\omegab}{\phib^2}\dot\phib\dot\varphi
	+\frac{2H}{\phib}\dot\varphi
	-\frac{2}{3\phib}\frac{\nabla^2}{a^2}\varphi
	-\frac{\dot\phib}{\phib}(\dot A+2\dot\psi)\nonumber\\
	&&-\left[\frac{\omegab\dot\phib^{2}}{\phib^2}+2\frac{\ddot\phib+2H\dot\phib}{\phib}\right]A\nonumber\\
	&&
	+\varphi\left[\left(\frac{\omegapb}{2\phib^2}-\frac{\omegab}{\phib^3}\right)\dot\phib^{2}
	-\frac{\ddot\phib+2H\dot\phib}{\phib^2}\right] \nonumber\\
	&&+\varphi\left[-\frac{\Vpb}{\phib}
	+\frac{\Vb}{\phib^2}\right],
	\label{eq:deltaPFinal}
\end{eqnarray}
which vanish in the pure transverse-traceless (TT) sector, as expected, since scalar quantities do not couple to tensor modes at linear order.

For the heat flux, the perturbation is defined as
\begin{equation}
	\delta q_i^{(\phi)}=-\delta T^{(\phi)}_{0i}-(\rhob+\Pb)\delta u_i,
	\label{eq:dqGeneral}
\end{equation}
where both the perturbation of the energy--momentum tensor and the velocity perturbation contribute. Evaluating this expression explicitly, one obtains
\begin{equation}
	{
		8\pi\,\delta q_i^{(\phi)}
		=
		-\frac{1}{\phib}\partial_i\dot\varphi
		+\frac{\dot\phib}{\phib}\partial_i A
		+\frac{\ddot\phib}{\phib\dot\phib}\partial_i\varphi.
	}
	\label{eq:dqExplicit}
\end{equation}
This form makes explicit the dependence on spatial gradients of the scalar perturbation and metric potential.

It is often convenient to rewrite this result in a more compact and physically transparent form. Using the definition of the velocity potential $v_\phi$ and the perturbed acceleration, one finds
\begin{equation}
	{
		8\pi\,\delta q_i^{(\phi)} = -\frac{|\dot\phib|}{\phib}\partial_i\left(A+\dot v_\phi\right) =- \frac{|\dot\phib|}{\phib}\delta a_i\,.
	}
	\label{eq:dqCompact}
\end{equation}
This expression highlights that the heat flux is directly proportional to the effective fluid acceleration, reinforcing the interpretation of the scalar field as an imperfect fluid with nontrivial transport properties.

This result shows that $q_a^{(\phi)}$ is purely scalar at linear order on a homogeneous FLRW background. In particular, it depends only on spatial gradients of scalar quantities, and therefore contains no transverse or tensor contributions. As a consequence, in the pure TT sector one has $\delta q_i^{(\phi)}=0$. Additionally, it exhibits the same structural form as the unperturbed $q_a^{(\phi)}$ found in the literature~\cite{Faraoni:2021lfc}, namely that the heat flux is proportional solely to the 4-acceleration, reinforcing the effective imperfect-fluid interpretation.

For the anisotropic stress, it is useful to distinguish between the trace-free (TF) scalar sector and the purely transverse-traceless (TT) tensor sector. The trace-free perturbed anisotropic stress is defined as
\begin{equation}
	\delta\pi^{(\phi)}_{ij}=
	\left[\delta T^{(\phi)}_{ij}-\Pb\,\delta g_{ij}-a^2\delta_{ij}\,\delta P_\phi\right]^{\TF},
	\label{eq:deltaPiProjection}
\end{equation}
where the subtraction of the trace ensures that only the anisotropic (shear-like) contributions are retained.

To make this decomposition explicit, introduce the scalar trace-free operator
\begin{equation}
	\Dij f \equiv \partial_i\partial_j f-\frac13\delta_{ij}\nabla^2 f,
	\label{eq:DijDef}
\end{equation}
which extracts the traceless part of second spatial derivatives of a scalar function. Using this operator, the scalar trace-free contribution to the anisotropic stress takes the form
\begin{equation}
	{
		8\pi\,\delta\pi_{ij}^{(\phi)}\big|_S
		=
		\frac{1}{\phib}\Dij \varphi,
	}
	\label{eq:piScalarFinal}
\end{equation}
showing that scalar perturbations generate anisotropic stress through spatial gradients of $\varphi$.

On the other hand, the pure TT contribution is given by
\begin{equation}
	{
		8\pi\,\delta\pi^{(\phi),\TT}_{ij}
		=
		-\frac{a^2}{2}\frac{\dot\phib}{\phib}\dot\chi_{ij}.
	}
	\label{eq:piTTMain}
\end{equation}
This term depends exclusively on the tensor perturbation $\chi_{ij}$ and its time derivative, and therefore captures the anisotropic stress associated with gravitational waves.

Thus, at linear order the scalar and tensor sectors activate distinct effective channels: the scalar sector sources both the heat-flux and scalar trace-free anisotropic-stress components, while the TT sector contributes only through the tensor anisotropic-stress channel. This clean separation reflects the decoupling of scalar and tensor modes in linear perturbation theory.

\section{Eckart connection and constitutive rewriting}\label{sec:EckartConnection}

The thermodynamic interpretation of scalar–tensor theories and their imperfect fluid component naturally follows from the fact that the corresponding field equations can be recast as effective Einstein equations sourced by an effective imperfect fluid, in addition to ordinary matter. This observation motivates an analysis of the thermodynamical framework associated with such a fluid, a direction that has indeed been explored in the literature. 

In this setting, Eckart’s first-order theory constitutes a natural starting point, since it accommodates the presence of a spacelike heat flux. Although this formulation is non-causal, it is nevertheless widely employed as a first approximation in relativistic thermodynamics.

\subsection{Eckart interpretation and effective shear viscosity}

In Eckart's first-order theory \cite{Eckart:1940te} the dissipative quantities such as the viscous pressure $P_{\text{vis}}$, heat current density $q_a$, and anisotropic stresses $\pi_{ab}$ are
related to the expansion $\Theta$, temperature $T$, and shear
tensor $\sigma_{ab}$ by the constitutive equations
\begin{equation}\label{eq:EckartHeat}
	q_a^{\text{Eckart}} = -\kappa\,h_a{}^b\left(\nabla_b T + T a_b\right), 
\end{equation}
\begin{equation}
	\pi_{ab}^{\text{Eckart}} = -2\eta\,\sigma_{ab}, \label{eq:EckartShear}
\end{equation}
\begin{equation}
	\Pi_{\rm bulk} = -\zeta\,\Theta, \label{eq:EckartBulk}
\end{equation}
where the shear tensor of the chosen congruence is defined as
\begin{equation}
	\sigma_{ab}=h_a{}^c h_b{}^d\nabla_{(c}u_{d)}-\frac13 h_{ab}\Theta,
	\label{eq:shear_def}
\end{equation}
and encodes the traceless, symmetric part of the velocity gradient orthogonal to $u^a$.

Since $\bar\sigma_{ij}=0$ on an FLRW background, the shear vanishes at zeroth order, and thus its perturbation directly captures first-order deviations. The linear spatial shear obeys
\begin{equation}
	\delta\sigma_{ij}=\left[\delta(\nabla_{(i}u_{j)})-H\,\delta g_{ij}\right]^{\TF},
	\label{eq:deltasigma_master2}
\end{equation}
which isolates the traceless part of the perturbed velocity gradients.

Using Eq.~\eqref{eq:duiBranch} together with the perturbed connection coefficients, one finds in the scalar sector
\begin{equation}
	{\delta\sigma_{ij}\big|_S=\Dij v_\phi,}
	\label{eq:sigma_scalar_final}
\end{equation}
while in the TT sector one obtains
\begin{equation}
	{\delta\sigma_{ij}\big|_{\TT}=\frac{a^2}{2}\dot\chi_{ij}.}
	\label{eq:sigma_TT_final}
\end{equation}
Thus, scalar perturbations contribute to the shear through spatial gradients of the velocity potential, whereas tensor perturbations contribute through the time evolution of gravitational waves.

Because $v_\phi=-\varphi/\dot\phib$, Eq.~\eqref{eq:sigma_scalar_final} implies
\begin{equation}
	\Dij\varphi=-\dot\phib\,\delta\sigma_{ij}\big|_S,
	\label{eq:Dphi_sigma_scalar}
\end{equation}
establishing a direct relation between scalar-field gradients and the shear tensor.

Comparing this result with Eqs.~\eqref{eq:piScalarFinal} and \eqref{eq:piTTMain}, one finds for the scalar and tensor sectors, respectively,
\begin{equation}
	8\pi\,\delta\pi_{ij}^{(\phi)}\big|_S=-\frac{\dot\phib}{\phib}\,\delta\sigma_{ij}\big|_S,
	\label{eq:pi_sigma_scalar}
\end{equation}
and
\begin{equation}
	8\pi\,\delta\pi_{ij}^{(\phi)}\big|_{\TT}=-\frac{\dot\phib}{\phib}\,\delta\sigma_{ij}\big|_{\TT}.
	\label{eq:pi_sigma_TT}
\end{equation}

Therefore, the same proportionality holds in both the scalar trace-free and TT sectors:
\begin{equation}
	{
		8\pi\,\delta\pi_{ij}^{(\phi)}=-\frac{\dot\phib}{\phib}\,\delta\sigma_{ij},
		\quad
		\text{(scalar tracefree and TT sectors).}
	}
	\label{eq:pi_sigma_combined}
\end{equation}
This unified relation highlights that the anisotropic stress is entirely determined by the shear of the effective fluid.

Equivalently, this can be written in the standard viscous form
\begin{equation}
	{
		\delta\pi_{ij}^{(\phi)}=-2\eta_{\rm eff}\,\delta\sigma_{ij},
		\qquad
		\eta_{\rm eff}=\frac{\dot\phib}{16\pi\phib},
	}
	\label{eq:pi_sigma_viscous_final}
\end{equation}
which matches exactly the structure of a first-order Eckart or Landau--Lifshitz shear-viscous constitutive relation~\cite{Faraoni:2021lfc}. This correspondence provides a clear physical interpretation: the scalar field behaves as an imperfect fluid with an effective shear viscosity determined dynamically by the background evolution of $\phi$.

\subsection{Perturbed Fourier law and matching to the effective scalar heat flux}

Furthermore, a central ingredient in the effective Eckart interpretation of scalar--tensor gravity is the generalized Fourier law
\begin{equation}
	q_a^{\rm Eckart}
	=
	-\kappa\left(D_aT+T\,a_a\right),
	\qquad
	D_aT\equiv h_a{}^b\nabla_bT,
	\label{eq:EckartFourierDaT}
\end{equation}
which relates the heat flux to both the projected temperature gradient and the fluid acceleration.

In the scalar-comoving frame of Jordan-like scalar--tensor gravity, it has been shown that
\begin{equation}
	q_a^{(\phi)} = q_a^{\text{Eckart}} 
\end{equation}
with
\begin{equation}\label{eq:TSTTgeneral}
	\kappa T = \frac{\sqrt{-\nabla_a\phi\nabla^a\phi}}{8\pi \phi}\,,
\end{equation}
and $D_a T = 0$. This last condition plays a crucial role in identifying the effective scalar heat flux with the Eckart description, as it enforces the vanishing of the spatially projected temperature gradient in that frame \cite{Faraoni:2021lfc}. However, more recent analyses have shown that this requirement can be overly restrictive: only the component of $D_aT$ orthogonal to the 4-acceleration is genuinely constrained, while a component parallel to $a_a$ can be absorbed into the acceleration term in Eq.~\eqref{eq:EckartFourierDaT} \cite{Pereira:2025dmk}.

We now compute $\delta q_{a}^{\text{Eckart}}$, translate these statements into first-order perturbation theory around FLRW, and determine under which conditions $\delta q_a^{(\phi)} = \delta q_{a}^{\text{Eckart}}$ holds.

We start by perturbing around spatially flat FLRW in cosmic time and write
\begin{equation}
	T=\bar T(t)+\delta T(t,\mathbf{x}),
	\qquad
	u^a=\bar u^a+\delta u^a.
\end{equation}
Because the background is homogeneous, one has
\begin{equation}
	\bar\nabla_a\bar T=(\dot{\bar T},0,0,0),
	\qquad
	\partial_i\bar T=0.
	\label{eq:bgTempGrad}
\end{equation}
so that only temporal gradients are present at zeroth order.

For scalar perturbations in the scalar-gradient frame, we use Eq.~\eqref{eq:duiBranch}. With this setup, the exact Eckart law \eqref{eq:EckartFourierDaT} yields, to first order around FLRW,
\begin{equation}
	\delta q_i^{\rm Eckart}
	=
	-\bar\kappa\left(\partial_i\delta T+\dot{\bar T}\,\delta u_i+\bar T\,\delta a_i\right).
	\label{eq:deltaqEckartGeneral}
\end{equation}
This expression clearly separates the contributions from spatial temperature gradients, velocity perturbations, and acceleration.

Comparing Eq.~\eqref{eq:deltaqEckartGeneral} with Eq.~\eqref{eq:dqCompact}, one finds that matching the effective scalar-field heat flux to an Eckart-type constitutive law requires
\begin{equation}
	\partial_i \delta T + \dot{\bar T}\,\delta u_i = 0.
	\label{eq:cond_projected_tempgrad}
\end{equation}
As will be discussed below, this condition is equivalent to the vanishing of the \emph{projected} temperature-gradient contribution at linear order. A sufficient (but stronger) specialization is to impose separately $\partial_i\delta T=0$ and $\dot{\bar T}\,\delta u_i=0$, for example by considering an isothermal background ($\dot{\bar T}=0$) and/or a comoving slicing.

Under the condition \eqref{eq:cond_projected_tempgrad}, the Eckart heat flux reduces to a purely acceleration-driven form, and one obtains
\begin{equation}\label{eq:Tback}
	\delta q_{i}^{(\phi)} = \delta q_{i}^{\text{Eckart}}
	\qquad\Longrightarrow\qquad
	\bar{\kappa}\,\bar{T} = \frac{|\dot{\bar{\phi}}|}{8\pi\,\bar{\phi}}\,,
\end{equation}
which coincides with the effective temperature identified in the unperturbed case~\cite{Faraoni:2021lfc}.

Although the condition \eqref{eq:cond_projected_tempgrad} may at first seem unrelated to the covariant requirement $D_aT=0$, it is in fact its precise perturbative realization. To see this explicitly, define the projected temperature gradient
\begin{equation}
	\mathcal{G}_a \equiv D_aT \equiv h_a{}^{b}\nabla_b T,
\end{equation}
and perturb it to first order around FLRW:
\begin{align}
	\delta\mathcal{G}_a
	&=
	\delta\!\left(h_a{}^{b}\nabla_bT\right)
	=
	\delta h_a{}^{b}\,\bar\nabla_b\bar T
	+\bar h_a{}^{b}\,\delta(\nabla_bT),
	\label{eq:deltaDaTstart}
\end{align}
where we used $\bar\nabla_b\bar T=\partial_b\bar T$ and $\delta(\nabla_bT)=\partial_b\delta T$, since $T$ is a scalar.

For spatial components ($a=i$) on FLRW one has $\bar h_i{}^{0}=0$ and $\bar h_i{}^{j}=\delta_i{}^{j}$. Moreover, using $h_i{}^{b}=\delta_i{}^{b}+u_i u^{b}$ together with $\bar u_i=0$ and $\bar u^{0}=1$, it follows that $\delta h_i{}^{0}=\delta u_i$ and $\delta h_i{}^{j}=0$. Equation~\eqref{eq:deltaDaTstart} then reduces to
\begin{align}
	\delta\mathcal{G}_i
	&=
	\delta h_i{}^{0}\,\dot{\bar T}
	+\bar h_i{}^{j}\,\partial_j\delta T
	=
	\dot{\bar T}\,\delta u_i+\partial_i\delta T.
	\label{eq:deltaDaTi}
\end{align}

Therefore, if the exact comoving-frame condition $D_aT=0$ is imposed, then $\delta\mathcal{G}_i=0$ and one obtains the correlated perturbative constraint
\begin{equation}
	\partial_i\delta T=-\dot{\bar T}\,\delta u_i.
	\label{eq:perturbedComovingTempCondition}
\end{equation}

In other words, the natural linear perturbative extension of $D_aT=0$ is \emph{not} $\partial_i\delta T=0$, but rather the relation \eqref{eq:perturbedComovingTempCondition} between the spatial temperature perturbation and the velocity perturbation. This condition ensures that the \emph{projected} temperature gradient vanishes in the perturbed fluid rest frame, providing the precise criterion for matching the scalar-field heat flux to the Eckart form at first order.

The other possibility for matching the heat fluxes is more subtle.  
Starting from the condition
\begin{equation}
\partial_i \delta T = 0,
\end{equation}
it follows that $\delta T$ carries no spatial gradients and therefore contains only a homogeneous ($k=0$) mode: the temperature perturbation is spatially uniform on the chosen time-slicing, i.e.\ it does not contain inhomogeneous ($k\neq 0$) fluctuations. Then to match both heat fluxes one must have
\begin{equation}
\dot{\bar T}\,\delta u_i=0.
\end{equation}

This admits two possibilities: $\dot{\bar T}=0$ or $\delta u_i=0$. The latter, in the scalar-gradient frame, is equivalent to $\partial_i\varphi=0$, and therefore imposes a strong restriction on the scalar perturbation sector, making it unsuitable for a generic perturbative treatment.

The first possibility, $\dot{\bar T}=0$, corresponds to an isothermal background.  
Using Eq.~\eqref{eq:Tback} and differentiating with respect to time, one finds
\begin{equation}
\dot{\bar\kappa}\,\bar T+\bar\kappa\,\dot{\bar T}
=
\frac{1}{8\pi}\frac{d}{dt}\!\left(\frac{|\dot{\bar\phi}|}{\bar\phi}\right),
\end{equation}
thus imposing $\dot{\bar T}=0$ yields the more general relation
\begin{equation}
\frac{d}{dt}\!\left(\frac{|\dot{\bar\phi}|}{\bar\phi}\right)
=
8\pi\,\dot{\bar\kappa}\,\bar T
=
\frac{\dot{\bar\kappa}}{\bar\kappa}\,
\frac{|\dot{\bar\phi}|}{\bar\phi},
\end{equation}
where in the last step we used again Eq.~\eqref{eq:Tback} and assumed $\bar\kappa\neq 0$. Therefore, on the isothermal branch $\dot{\bar T}=0$, one has
\begin{equation}
\frac{|\dot{\bar\phi}|}{\bar\phi}=C\,\bar\kappa(t),
\qquad
C\equiv 8\pi \bar T=\text{const.},
\label{eq:generalized_scaling_condition}
\end{equation}
where $C$ is a positive constant. Assuming a monotonic background scalar so that
$\mathrm{sgn}(\dot{\bar\phi})$ is fixed, Eq.~\eqref{eq:generalized_scaling_condition}
can be rewritten as
\begin{equation}
\frac{\dot{\bar\phi}}{\bar\phi}=\sigma\,C\,\bar\kappa(t),
\qquad
\sigma\equiv \mathrm{sgn}(\dot{\bar\phi})=\pm1,
\end{equation}
which integrates immediately to
\begin{equation}
\bar\phi(t)=\phi_0\,
\exp\!\left[\sigma C\int^t \bar\kappa(t')\,dt'\right].
\label{eq:generalized_scaling_solution}
\end{equation}

Hence, in the general isothermal case the background scalar is not necessarily a simple
exponential in cosmic time. Rather, it satisfies the generalized scaling relation
\eqref{eq:generalized_scaling_solution}, determined by the conductivity profile
$\bar\kappa(t)$. The familiar exponential (self-similar/scaling) subclass is recovered
when the conductivity is also time-independent, $\dot{\bar\kappa}=0$, in which case
\begin{equation}
\frac{|\dot{\bar\phi}|}{\bar\phi}=\text{const.}
\qquad\Longrightarrow\qquad
\bar\phi(t)=\phi_0\,e^{Ct}.
\end{equation}

These exponential scaling backgrounds are of particular interest because they are self-similar solutions that often act as late-time attractors and permit analytic control of the cosmological dynamics in scalar--tensor theories and related scalar-field models \cite{CopelandLiddleWands1998ExponentialScaling,NunesMimoso2000ScalingPotentials,BarrowMimoso1994PerfectFluidSTT,MimosoNunes1998GRAttractor,NunesMimosoCharters2001InteractingFluids}. More generally, however, in the nontrivial scalar branch selected by $\partial_i\delta T=0$ together with exact flux matching, one obtains the broader compatibility class of backgrounds obeying the generalized scaling law \eqref{eq:generalized_scaling_solution}, with the exponential subclass arising when $\bar\kappa$ is constant. 

Moreover, it is important to stress the status of this result. Since the gravitational sector fixes the invariant product $\bar\kappa\bar T$, rather than $\bar\kappa$ and $\bar T$ separately, Eq.~\eqref{eq:generalized_scaling_solution} should be interpreted as a constitutive compatibility relation between the isothermal choice and the conductivity profile, not as a restriction derived from the scalar--tensor field equations alone. Once $\bar\kappa(t)$ is specified independently, however, it becomes a genuine condition on the allowed background scalar evolution.

Furthermore, imposing $\partial_i\delta T=0$ in Newtonian gauge removes the gradient-driven conductive contribution to the Eckart flux at linear order, so that the latter reduces to
\begin{equation}
\delta q_i^{\rm Eck}
=
-\bar\kappa\left(\dot{\bar T}\,\delta u_i+\bar T\,\delta a_i\right).
\end{equation}
This can be matched directly to the scalar-sector result $\delta q_i^{(\phi)}\propto \delta a_i$ provided $\dot{\bar T}\,\delta u_i=0$. For the nontrivial scalar sector this leads naturally to the isothermal branch $\dot{\bar T}=0$, which in turn constrains the background evolution through the generalized scaling law above. In this sense, the seemingly mild assumption of vanishing inhomogeneous temperature perturbations singles out, through flux matching, a distinguished class of scalar--tensor backgrounds, whose constant-conductivity limit reproduces the well-known exponential scaling sector.

The previous discussion can be sharpened using the decomposition introduced in \cite{Pereira:2025dmk}. The key point is that one should not require the entire spatial temperature gradient $D_aT$ to vanish. Rather, one should decompose it into a component parallel to the acceleration and a component orthogonal to it.

At first order in the scalar sector, define
\begin{equation}
\mathcal T_i\equiv \partial_i\delta T+\dot{\bar T}\,\delta u_i.
\label{eq:GiDef}
\end{equation}

This is precisely the perturbation of the spatially projected temperature gradient:
\begin{equation}
\delta(D_iT)=\mathcal T_i.
\end{equation}
We may decompose $\mathcal T_i$ as
\begin{equation}
{
\mathcal T_i=\alpha\,\delta a_i+\Xi_i^{\perp},
\qquad
\Xi_i^{\perp}\,\delta a^i=0,
}
\label{eq:GiDecomposition}
\end{equation}
where $\alpha$ is a scalar coefficient and $\Xi_i^\perp$ is the component orthogonal to $\delta a_i$.

Substituting \eqref{eq:GiDecomposition} into \eqref{eq:deltaqEckartGeneral} gives
\begin{equation}
{
\delta q_i^{\rm Eckart}
=
-\bar\kappa\Big[(\bar T+\alpha)\,\delta a_i+\Xi_i^\perp\Big],
}
\label{eq:deltaqEckartDecomposed}
\end{equation}
that makes the interpretation transparent: the piece of $D_iT$ parallel to $\delta a_i$ is \emph{not} an obstruction to an Eckart-like description; it can be absorbed into an effective coefficient multiplying $\delta a_i$, only the orthogonal piece $\Xi_i^\perp$ is a genuine obstruction to the standard Eckart form.

The scalar--tensor effective heat flux is purely aligned with $\delta a_i$. Therefore an exact Eckart matching requires
\begin{equation}
{
\Xi_i^\perp=0,
}
\label{eq:XiPerpZeroCondition}
\end{equation}
and then the parallel coefficient must satisfy
\begin{equation}
{
\bar\kappa(\bar T+\alpha)
=
-\frac{\dot{\bar\phi}}{8\pi\bar\phi},
}
\label{eq:CoeffMatchingParallel}
\end{equation}
up to the sign convention associated with the chosen orientation of $u^a$.

For linear scalar perturbations on FLRW, all scalar spatial vectors are gradients. In Fourier space, each scalar vector is proportional to $k_i$, and therefore, mode by mode,
\begin{equation}
\delta u_i\propto k_i,
\qquad
\partial_i\delta T\propto k_i,
\qquad
\delta a_i\propto k_i.
\end{equation}

Accordingly, there is no independent transverse scalar direction for a single Fourier mode, and the orthogonal piece in \eqref{eq:GiDecomposition} vanishes mode by mode:
\begin{equation}
{
\Xi_i^\perp=0
\qquad
\text{(scalar sector, mode by mode on FLRW).}
}
\label{eq:XiPerpZeroFLRWscalar}
\end{equation}

In this sense, the FLRW scalar sector belongs precisely to the class of highly symmetric situations discussed in \cite{Pereira:2025dmk}, where an Eckart-like description can survive even though the stronger condition $D_aT=0$ is unnecessary.

Having $\bar{\kappa}\bar{T}$ established allows also to rewrite $8\pi\delta \pi_{ij}^{(\phi)} = -\frac{\dot{\bar{\phi}}}{\bar{\phi}}\delta \sigma_{ij}$ as
\begin{equation}
    \delta \pi_{ij}^{(\phi)} = \bar{\kappa}\bar{T}\delta \sigma_{ij}
\end{equation}
and the Eckart connection
\begin{equation}\label{eq:pi_sigma_viscous_finalT}
{
\delta\pi_{ij}^{(\phi)}=-2\eta_{\rm eff}\,\delta\sigma_{ij},
\qquad
\eta_{\rm eff}=-\frac{\bar{\kappa}{\bar{T}}}{2}<0,
}
\end{equation}
that is equal to the $\eta$ obtained in the non-perturbative case~\cite{Faraoni:2021lfc}.

\subsection{Entropy density and its perturbation}

Following the Eckart-inspired treatment of the effective $\phi$-fluid, we introduce an entropy
density (per physical volume) via the first-law expression~\cite{Faraoni:2021lfc,Faraoni:2021jri}
\begin{equation}
s \equiv \frac{\rho+P}{T},
\label{eq:entropy_density_def}
\end{equation}
where $(\rho,P)$ are the effective energy density and isotropic pressure of the sector under
consideration and $T$ is the corresponding (effective) temperature.

Splitting into background plus perturbation,
\begin{equation}
\rho=\bar\rho+\delta\rho,\qquad P=\bar P+\delta P,\qquad T=\bar T+\delta T,
\end{equation}
and expanding \eqref{eq:entropy_density_def} to first order yields
\begin{equation}
{
\delta s
=
\frac{\delta\rho+\delta P}{\bar T}
-\frac{\bar\rho+\bar P}{\bar T^{\,2}}\;\delta T.
}
\label{eq:delta_s}
\end{equation}

In Eckart's formalism the entropy current associated with heat transport is $q^a/T$, and the total entropy current can be written as~\cite{Eckart:1940te,Maartens:1996vi}
\begin{equation}
{
s^a \equiv s\,u^a + \frac{q^a}{T},
\qquad
u^a u_a=-1,\qquad u_a q^a=0.
}
\label{eq:entropy_current_def}
\end{equation}

On an FLRW background in the comoving frame one has $\bar q^a=0$, hence $\bar s^a=\bar s\,\bar u^a$.

To first order (and using $\bar q^a=0$), the perturbation of \eqref{eq:entropy_current_def} is
\begin{equation}
{
\delta s^a = \delta s\,\bar u^a + \bar s\,\delta u^a + \frac{1}{\bar T}\,\delta q^a,
}
\label{eq:delta_sa}
\end{equation}
where we neglected the term $-(\bar q^a/\bar T^2)\delta T$ because $\bar q^a=0$.
In Newtonian gauge (no shift) one has $\delta q^0=0$ at linear order (since $q^a$ is spatial),
and $\delta u^0=-A$, so
\begin{equation}
\delta s^0 = \delta s - \bar s\,A,
\qquad
\delta s^i = \bar s\,\delta u^i + \frac{1}{\bar T}\,\delta q^i.
\label{eq:delta_sa_components}
\end{equation}

For a particle-conserving Eckart fluid, the divergence of the entropy current can be expressed entirely in terms of the dissipative quantities (bulk viscous pressure $P_{\rm vis}$, heat flux $q^a$, and anisotropic stress $\pi_{ab}$) as~\cite{Eckart:1940te,Maartens:1996vi}
\begin{equation}
\nabla_a s^a
=
\frac{\Pi_{\rm bulk}^2}{\zeta\,T}
+
\frac{q_a q^a}{\kappa\,T^2}
+
\frac{\pi_{ab}\pi^{ab}}{2\eta\,T},
\label{eq:entropy_production_quadratic}
\end{equation}
with transport coefficients $(\zeta,\kappa,\eta)$. For the effective $\phi$-fluid the bulk term is absent ($\zeta=0$), and in the comoving frame the heat flux is purely ``inertial'' (acceleration-driven), while the anisotropic stress is shear-driven.

On an FLRW background one has $\bar a_i=0$ and $\bar\sigma_{ij}=0$, hence
\begin{equation}
\overline{\nabla_a s^a}=0.
\label{eq:background_entropy_prod_zero}
\end{equation}

Moreover, since \eqref{eq:entropy_production_quadratic} is \emph{quadratic} in the dissipative
fluxes and stresses, it follows immediately that its \emph{linear} perturbation vanishes:
\begin{equation}
{
\delta(\nabla_a s^a)=0
\qquad\text{(about FLRW at first order).}
}
\label{eq:delta_entropy_prod_zero}
\end{equation}

Therefore, entropy production (or decrease) appears only at second order in perturbations around FLRW. The first non-trivial contribution is the second-order piece obtained by inserting the first-order dissipative quantities into \eqref{eq:entropy_production_quadratic}
\begin{equation}
{
\left(\nabla_a s^a\right)^{(2)}
=
\frac{\delta q_a\,\delta q^a}{\bar\kappa\,\bar T^{\,2}}
+
\frac{\delta\pi_{ab}\,\delta\pi^{ab}}{2\bar\eta\,\bar T}.
}
\label{eq:entropy_prod_second_order_general}
\end{equation}

In the $\phi$-comoving frame used in the thermodynamic interpretation, the constitutive
relations reduce to $q_a=-(\kappa T)a_a$ and $\eta=-(\kappa T)/2$, implying
\begin{equation}
{
\left(\nabla_a s^a\right)^{(2)}
=
\bar\kappa\left(\delta a_a\,\delta a^a - \delta\sigma_{ab}\,\delta\sigma^{ab}\right).
}
\label{eq:entropy_prod_second_order_phi}
\end{equation}

In particular, in the pure TT sector $\delta a_i=0$ while $\delta\sigma_{ij}\neq 0$, so the shear
term contributes with a negative sign, reflecting the negative effective shear viscosity of the
$\phi$-fluid. This is consistent with the fact that the effective $\phi$-system is not isolated and its entropy need not be monotonic. This result is conceptually important for the interpretation of the effective thermal picture. On an FLRW background, the linear theory already exhibits the full constitutive structure of the effective $\phi$ sector---effective temperature, heat flux, anisotropic stress, viscosity, and entropy current---but the associated entropy production vanishes identically at first order because it is quadratic in the dissipative variables. Thus, the linear analysis captures the transport-like \emph{kinematics} of the effective description without yet probing its genuinely irreversible content.

For this reason, the absence of entropy production at first order should not be viewed as evidence against a possible thermal interpretation of scalar--tensor gravitational perturbations. Rather, it shows that the first nontrivial thermodynamic diagnostics necessarily arise only at second order, where products of the first-order heat-flux and anisotropic-stress channels can contribute. In particular, for the tensor sector, this indicates that any sharper assessment of whether scalar--tensor gravitational waves possess deeper thermal properties must rely on a second-order perturbative analysis.

At the same time, because the effective $\phi$ sector is not an ordinary isolated material fluid and may carry a negative effective shear viscosity, the second-order entropy balance need not satisfy the positivity properties familiar from standard irreversible thermodynamics. This reinforces the interpretation of the present construction as an effective constitutive framework, while also identifying second-order perturbation theory as the natural arena in which to test how far the thermal analogy can be pushed.

\section{Perturbed heating/cooling law for $\kappa T$}\label{sec:deltaKTevolution}

\subsection{Linearized evolution of the effective scalar temperature}

In the thermal interpretation of scalar--tensor gravity, one introduces the (non-negative) scalar~\eqref{eq:TSTTgeneral} and constructs an exact ``heating/cooling'' equation for $\kappa T$ of the form%
\footnote{Ref.~\cite{Faraoni:2025alq} provides a general scalar--tensor evolution equation and its equivalent rewriting in terms of $\kappa T$, the expansion $\Theta\equiv\nabla_a u^a$, the matter trace, and the potential. We denote by $\omega(\phi)$ and $V(\phi)$ the usual scalar--tensor functions.}
\begin{eqnarray}\label{eq:KT_evol_general}
	\frac{d(\kappa T)}{d\tau} &=& 8\pi\,\chi(\phi)\,(\kappa T)^2-\Theta\,\kappa T +\frac{T^{(m)}}{(2\omega+3)\,\phi} 		
	\nonumber\\
	&&+\frac{V_{,\phi}-2V/\phi}{4\pi\,(2\omega+3)},
\end{eqnarray}
where
\begin{equation}
	\chi(\phi)\equiv
	1+\frac{\phi}{2}\,\frac{d}{d\phi}\ln(2\omega+3),
	\label{eq:Xi_def}
\end{equation}
and $d/d\tau\equiv u^a\nabla_a$, $T^{(m)}\equiv T^{(m)a}{}_{a}$.%
\footnote{For electrovacuum or conformal matter, $T^{(m)}=0$. In Brans--Dicke theory with $\omega=\mathrm{const.}$ and $V=0$, Eq.~\eqref{eq:KT_evol_general} reduces to $d(\kappa T)/d\tau = \kappa T(8\pi \kappa T-\Theta)$, as discussed in Ref.~\cite{Faraoni:2025alq}.}
The explicit Brans--Dicke reduction and more general cases are given in Ref.~\cite{Faraoni:2025alq}.

Using the perturbed operator $\delta (\nabla\phi)^2$ (see \ref{app:operators}), one has
\begin{equation}
	\delta X = 2\dot{\bar\phi}\,(\dot\varphi-\dot{\bar\phi}A),
	\qquad
	\delta\sqrt{X}
	=
	\frac{\delta X}{2\sqrt{\bar X}}
	=
	\frac{\dot{\bar\phi}}{|\dot{\bar\phi}|}\,(\dot\varphi-\dot{\bar\phi}A),
	\label{eq:drootX}
\end{equation}
leading to the linearized perturbation of the effective temperature,
\begin{equation}
	\delta(\kappa T)
	=
	\frac{1}{8\pi}\left[
	\frac{1}{\bar\phi}\,\frac{\dot{\bar\phi}}{|\dot{\bar\phi}|}\,(\dot\varphi-\dot{\bar\phi}A)
	-\frac{|\dot{\bar\phi}|}{\bar\phi^{\,2}}\,\varphi
	\right].
	\label{eq:dKT_general}
\end{equation}
For $\dot{\bar\phi}<0$, one may remove the absolute values using $|\dot{\bar\phi}|=-\dot{\bar\phi}$ and $\dot{\bar\phi}/|\dot{\bar\phi}|=-1$ if desired.

Since $\kappa T$ is a scalar, its convective derivative reads
\begin{equation}
	\frac{d(\kappa T)}{d\tau} = u^a\nabla_a (\kappa T) = u^a\partial_a (\kappa T).
\end{equation}
Linearizing about FLRW and using $\partial_i\overline{ \kappa T}=0$ gives
\begin{align}
	\delta\!\left(\frac{d(\kappa T)}{d\tau}\right)
	&=
	\delta u^0\,\dot{\overline{\kappa T}} + \dot{\delta(\kappa T)},
	\label{eq:delta_dKT_dtau_step}
\end{align}
where the normalization $u^a u_a=-1$ implies $\delta u^0=-A$, so that
\begin{equation}
	\delta\!\left(\frac{d(\kappa T)}{d\tau}\right)
	=
	\dot{\delta(\kappa T)}-A\,\dot{\overline{\kappa T}}.
	\label{eq:delta_dKT_dtau}
\end{equation}

Equation~\eqref{eq:KT_evol_general} can be compactly written as
\begin{equation}
	\frac{d(\kappa T)}{d\tau} = \mathcal{F}(\kappa T,\Theta,\phi,T^{(m)}),
\end{equation}
where $\mathcal{F}$ is given by the right-hand side of \eqref{eq:KT_evol_general}. Linearizing then yields
\begin{eqnarray}
	\delta\mathcal{F}
	&=& 16\pi\,\bar\chi\,\overline{\kappa T}\,\delta(\kappa T)
	+ 8\pi\,\overline{(\kappa T)}^{\,2}\,\delta\chi 
		\nonumber \\
	&&
	- \overline{\kappa T}\,\delta\Theta
	- \bar\Theta\,\delta(\kappa T) 
	+ \delta\mathcal{S},
	\label{eq:deltaF_general}
\end{eqnarray}
where $\bar\chi\equiv\chi(\bar\phi)$, 
\begin{equation}
	\delta\chi = \chi_{,\phi}(\bar\phi)\,\varphi,
\end{equation}
and the source perturbation is
\begin{equation}
	\delta\mathcal{S}\equiv
	\delta\!\left(\frac{T^{(m)}}{(2\omega+3)\phi}\right)
	+
	\delta\!\left(\frac{V_{,\phi}-2V/\phi}{4\pi(2\omega+3)}\right).
	\label{eq:delta_sources_general}
\end{equation}

Combining Eqs.~\eqref{eq:delta_dKT_dtau}, \eqref{eq:deltaF_general}, and \eqref{eq:deltaTheta} leads to the linearized evolution equation for the effective scalar temperature:
\begin{eqnarray}
	\dot{\delta(\kappa T)}-A\,\dot{\overline{\kappa T}} &=&
	\big(16\pi\,\bar\chi\,\overline{\kappa T}-\bar\Theta\big)\,\delta(\kappa T) 
	\nonumber \\
	&& + 8\pi\,\overline{(\kappa T)}^{\,2}\,\delta\chi
	- \overline{\kappa T}\,\delta\Theta
	+ \delta\mathcal{S}.
	\label{eq:deltaKT_evol_general}
\end{eqnarray}

In the electrovacuum Brans--Dicke limit, where $\omega=\mathrm{const.}$, $V=0$, and $T^{(m)}=0$, one has $\chi=1$ and $\delta\chi=\delta\mathcal{S}=0$. In this case, Eq.~\eqref{eq:deltaKT_evol_general} reduces to the simpler form
\begin{equation}
	\dot{\delta(\kappa T)}-A\,\dot{\overline{\kappa T}}
	=
	\big(16\pi\,\overline{\kappa T}-\bar\Theta\big)\,\delta(\kappa T)
	- \overline{\kappa T}\,\delta\Theta,
	\label{eq:deltaKT_evol_BD}
\end{equation}
which makes explicit the linearized dynamics of $\delta(\kappa T)$ in a Brans--Dicke electrovacuum background.

The background equation \eqref{eq:KT_evol_general} governs only $\overline{\kappa T}(t)$ and therefore only the homogeneous (``$k=0$'') thermal evolution.
By contrast, the perturbed law \eqref{eq:deltaKT_evol_general} provides genuinely new information. One can analyze the \emph{dynamics of inhomogeneities}. It determines whether $\delta(\kappa T)$ modes grow or decay, i.e.\ the linear stability of an isothermal (or quasi-isothermal) background in the thermal variable.

Moreover, it also provides information about the \emph{kinematical sourcing}. Even with the constitutive law $q^{(\phi)}_a=-\kappa T\,a_a$ fixing the \emph{background} coefficient $\overline{\kappa T}$, Eq.~\eqref{eq:deltaKT_evol_general} shows explicitly how $\delta\Theta$ and metric perturbations (through the $A\,\dot{\overline{\kappa T}}$ term) source $\delta(\kappa T)$. There is also a \emph{Non-trivial content beyond flux matching} because $\bar a_i=0$ on FLRW, the linearized flux law gives $\delta q^{(\phi)}_i=-\overline{\kappa T}\,\delta a_i$ independently of $\delta(\kappa T)$; thus matching perturbed heat fluxes constrains only $\overline{\kappa T}$. Equation \eqref{eq:deltaKT_evol_general} is the additional dynamical input that governs $\delta(\kappa T)$ itself. Furthermore, if one imposes a restriction such as $\delta(D_i T)=0$ (no conductive gradient in the rest space) or a gauge-fixed condition like $\partial_i\delta T=0$, Eq.~\eqref{eq:deltaKT_evol_general} provides the criterion for whether that restriction is preserved under time evolution or is regenerated by expansion/metric sources.

At linear order, $\delta(\kappa T)$ measures fluctuations in both the \emph{rate of change} of $\phi$ and the \emph{effective Planck mass} (or equivalently the effective gravitational coupling) in a way that necessarily involves metric perturbations through the perturbed lapse.
One finds, neglecting $(\partial_i\varphi)^2$ terms,
\begin{equation}
{
\frac{\delta(\kappa T)}{\overline{\kappa T}}
=
\frac{\dot\varphi}{\dot{\bar\phi}}-A-\frac{\varphi}{\bar\phi},
\qquad (\dot{\bar\phi}\neq 0),
}
\label{eq:dKT_fractional}
\end{equation}
where $\phi=\bar\phi+\varphi$ and $g_{00}=-(1+2A)$.
The three contributions in \eqref{eq:dKT_fractional} have a clear meaning:
(i) $\dot\varphi/\dot{\bar\phi}$ is the fractional perturbation of the scalar roll rate,
(ii) $-A$ corrects the conversion between coordinate time and perturbed proper time (a lapse effect), and
(iii) $-\varphi/\bar\phi$ is the fractional perturbation of the effective Planck mass in Jordan-frame scalar--tensor theory, where $M_*^2\propto\phi$.
Equivalently, since $G_{\rm eff}\sim 1/\phi$ in the Jordan frame, $\varphi/\bar\phi$ corresponds (up to sign) to $\delta G_{\rm eff}/G_{\rm eff}$.

A useful alternative form arises in the scalar-gradient frame, where $\delta u_i=\partial_i v_\phi$ with
$v_\phi\equiv-\varphi/\dot{\bar\phi}$, yielding
\begin{equation}
\delta(\kappa T)
=
\frac{\dot{\bar\phi}}{8\pi\bar\phi}\left[
A+\dot v_\phi+\left(\frac{\ddot{\bar\phi}}{\dot{\bar\phi}}-\frac{\dot{\bar\phi}}{\bar\phi}\right)v_\phi
\right],
\label{eq:dKT_in_vphi}
\end{equation}
which makes explicit that $\delta(\kappa T)$ contains the same kinematical combination
$A+\dot v_\phi$ that controls the acceleration perturbation $\delta a_i=\partial_i(A+\dot v_\phi)$ and hence the heat flux, \emph{but} also carries additional information through the $v_\phi$-dependent amplitude term governed by background rates.
Therefore, even when flux matching fixes $\overline{\kappa T}$ through \eqref{eq:dq_onlyKTbar}, the perturbation $\delta(\kappa T)$ retains independent content and must be determined by the perturbed heating/cooling law (or, equivalently, by the full linearized scalar--tensor field equations).

\subsection{Gauge behaviour and the gauge-invariant projected gradient}\label{subsec:dKT_gauge}

Since $\kappa T$ is a scalar, its Lie derivative along a vector field $\xi^a$ gives $\pounds_\xi\overline{\kappa T}=\xi^a\bar\nabla_a\overline{\kappa T}=\xi^0\,\dot{\overline{\kappa T}}$. Consequently, under a first-order gauge transformation, one has
\begin{equation}
	\delta(\kappa T)\ \longrightarrow\ \delta(\kappa T)' = \delta(\kappa T) + \dot{\overline{\kappa T}}\,\xi^0,
	\label{eq:dKT_gauge_transf}
\end{equation}
so that $\delta(\kappa T)$ is generally gauge dependent whenever $\dot{\overline{\kappa T}}\neq 0$. In particular, for spatial derivatives,
\begin{equation}
	\partial_i \delta(\kappa T)' = \partial_i \delta(\kappa T) + \dot{\overline{\kappa T}}\,\partial_i \xi^0,
	\label{eq:grad_dKT_gauge}
\end{equation}
demonstrating that conditions such as $\partial_i \delta(\kappa T)=0$ are only meaningful once a perturbative gauge has been fixed.

The covariant quantity controlling heat conduction in an Eckart-type framework is the \emph{spatially projected gradient}
\begin{equation}
	D_a(\kappa T) \equiv h_a{}^{b}\nabla_b(\kappa T).
	\label{eq:proj_grad_KT}
\end{equation}

On a FLRW background, $\overline{D_a(\kappa T)}=0$. Therefore, by the Stewart--Walker lemma~\cite{Stewart:1974uz}, the linear perturbation $\delta(D_a(\kappa T))$ is gauge invariant, and in particular its spatial components are also gauge invariant. Explicitly, in the scalar sector one finds
\begin{equation}
	\delta\!\big(D_i(\kappa T)\big)
	=
	\partial_i \delta(\kappa T) + \dot{\overline{\kappa T}}\,\delta u_i
	=
	\partial_i \!\big(\delta(\kappa T) + \dot{\overline{\kappa T}}\,v_\phi \big),
	\label{eq:dDiKT}
\end{equation}
which naturally identifies the gauge-invariant rest-frame perturbation
\begin{equation}
	\delta(\kappa T)_{\rm rf} \equiv \delta(\kappa T) + \dot{\overline{\kappa T}}\,v_\phi,
	\label{eq:dKT_restframe}
\end{equation}
for which $\delta(D_i(\kappa T)) = \partial_i \delta(\kappa T)_{\rm rf}$. Moreover, in backgrounds with constant $\overline{\kappa T}$, the gauge ambiguity in \eqref{eq:dKT_gauge_transf} disappears, so that $\delta(\kappa T)$ itself provides a gauge-invariant characterization of thermal inhomogeneities. Regardless, the physically relevant measure of conduction remains the gauge-invariant projected gradient $\delta(D_i(\kappa T))$.


\subsection{Degeneracy of $\delta(\kappa T)$ from FLRW heat-flux matching}\label{subsec:dKT_degeneracy}

The thermal constitutive law for the effective scalar heat flux takes the acceleration-driven form
\begin{equation}
	q^{(\phi)}_a = -\,\kappa T\,a_a,
	\label{eq:q_accel_repeat}
\end{equation}
which implies, upon perturbation,
\begin{equation}
	\delta q_i^{(\phi)} = -\,\overline{\kappa T}\,\delta a_i - \delta(\kappa T)\,\bar a_i.
	\label{eq:dq_split}
\end{equation}

However, for homogeneous FLRW backgrounds the effective flow is geodesic and therefore
\begin{equation}
	\bar a_i=0,
	\label{eq:aibar_zero}
\end{equation}
so that the linearized constitutive law collapses to
\begin{equation}
	{
		\delta q_i^{(\phi)} = -\,\overline{\kappa T}\,\delta a_i,
	}
	\label{eq:dq_onlyKTbar}
\end{equation}
independently of $\delta(\kappa T)$. This structural degeneracy explains why matching perturbed heat fluxes about FLRW provides a powerful \emph{consistency check}, i.e., it reproduces the same background coefficient $\overline{\kappa T}$ obtained in the homogeneous analysis, but cannot by itself constrain $\delta(\kappa T)$.

The perturbation $\delta(\kappa T)$ becomes observable through flux relations only beyond this restricted setting, e.g.\ on non-FLRW (non-geodesic) backgrounds where $\bar a_i\neq 0$, at second order around FLRW where products of first-order quantities generate terms $\sim \delta(\kappa T)\,\delta a_i$, or, most directly, by invoking the (perturbed) heating/cooling equation, which supplies an independent dynamical evolution law for $\delta(\kappa T)$.

\section{Einstein-like perturbation equations and thermal channels}\label{sec:EinsteinSideChannels}

The identifications derived above between Eckart's dissipative quantities and the corresponding effective $\phi$-sector projections can be inserted directly into the perturbed Einstein-like and scalar-field equations. This makes it possible to interpret several features of the perturbed system in effective thermodynamic terms.

We will start by decomposing the total Einstein-like source as
\begin{equation}
\Tmath_{ab}=\rho_{\rm E}u_a u_b + P_{\rm E}h_{ab}+2u_{(a}q^{\rm(E)}_{b)}+\pi^{\rm(E)}_{ab}.
\label{eq:EinSideFluidDecomp}
\end{equation}

At first order on FLRW the projected perturbations split as
\begin{align}
\delta\rho_{\rm E}
&=
\frac{1}{\phib}\,\delta\rho_m
-\frac{\varphi}{\phib^2}\,\bar\rho_m
+\delta\rho_\phi,
\label{eq:rhoEinSplitExplicit}\\
\delta P_{\rm E}
&=
\frac{1}{\phib}\,\delta P_m
-\frac{\varphi}{\phib^2}\,\bar P_m
+\delta P_\phi,
\label{eq:PEinSplitExplicit}\\
\delta q_i^{\rm(E)}
&=
\frac{1}{\phib}\,\delta q_i^{(m)}+\delta q_i^{(\phi)},
\label{eq:qEinSplitExplicit}\\
\delta\pi_{ij}^{\rm(E)}
&=
\frac{1}{\phib}\,\delta\pi_{ij}^{(m)}+\delta\pi_{ij}^{(\phi)},
\label{eq:piEinSplitExplicit}
\end{align}
where $\bar q_i^{(m)}=\bar\pi_{ij}^{(m)}=0$ because the background matter source is isotropic with respect to the FLRW/scalar-gradient congruence $\bar u^a=(1,0,0,0)$, whereas $\bar\rho_m$ and $\bar P_m$ need not vanish.

\subsection{Scalar sector}\label{sec:scalarpart}

In the Newtonian gauge and for scalar perturbations, the linearized Einstein-tensor components on a spatially flat FLRW background are standard and can be found, in equivalent sign and index conventions, in Refs.~\cite{Bardeen:1980kt,Kodama:1984ziu,Mukhanov:1990me,Ma:1994dv,Hwang:1996xh,Malik:2008im}. Since in the present work the matter and effective-fluid channels are written with covariant components, it is convenient to display the Einstein tensor in the same form:
\begin{align}
\delta G_{00}
&=-2\left[3H(\dot\psi+HA)-\frac{\nabla^2}{a^2}\psi\right]+6H^2A,
\label{eq:dG00ScalarCov}\\
\delta G_{0i}\big|_S
&=2\,\partial_i(\dot\psi+HA),
\label{eq:dG0iScalarCov}\\
\left(\delta G_{ij}\right)_{\TF,S}
&=-\Dij(A-\psi),
\label{eq:dGijTFScalarCov}\\
\frac{1}{3a^2}\,\delta^{ij}\delta G_{ij}
&=-2\left[\tilde{H}-\frac{1}{3a^2}\nabla^2(A-\psi)\right].
\label{eq:dGtraceScalarCov}
\end{align}
where
\begin{equation}
 \tilde{H}=  \ddot\psi+H(\dot A+3\dot\psi)+
(2\dot H+3H^2)(A-\psi)\,.
\end{equation}

The first line follows from lowering the index on the standard mixed component and adding the background correction $\delta g_{00}\,\bar G^0{}_0$, while the scalar trace and tracefree spatial pieces follow from the usual decomposition of $\delta G_{ij}$ into its isotropic and tracefree parts on the Euclidean spatial slices \cite{Kodama:1984ziu,Mukhanov:1990me,Ma:1994dv,Malik:2008im}. To connect these geometric expressions with the effective thermodynamic channels, we now project the Einstein-like equation~\eqref{eq:EinsteinLikeTmathfrak} onto the observer congruence. Using
\begin{equation}
\delta\rho_{\rm E}=\delta T^{\rm(E)}_{00}-2\bar\rho_{\rm E}A,
\qquad
\delta P_{\rm E}=\frac{1}{3a^2}\delta^{ij}\delta T^{\rm(E)}_{ij}+2\bar P_{\rm E}\psi,
\label{eq:EinSideProjectedDefs}
\end{equation}
and $\bar q_i^{\rm(E)}=0$, $\bar\pi_{ij}^{\rm(E)}=0$, together with the background Einstein equations $3H^2=8\pi\bar\rho_{\rm E}$ and $-(2\dot H+3H^2)=8\pi\bar P_{\rm E}$, Eqs.~\eqref{eq:dG00ScalarCov}--\eqref{eq:dGtraceScalarCov} become
\begin{align}
3H(\dot\psi+HA)-\frac{\nabla^2}{a^2}\psi
&=-4\pi\,\delta\rho_{\rm E},
\label{eq:ScalarHamEinSide}\\
\partial_i(\dot\psi+HA-\dot Hv_\phi)&=-4\pi\delta q_i^{\rm(E)},
\label{eq:ScalarMomEinSide}\\
\tilde{H}+(2\dot H+3H^2)\psi-\frac{1}{3a^2}\nabla^2(A-\psi)
&=-4\pi\delta P_{\rm E},
\label{eq:ScalarPressEinSide}\\
-\Dij(A-\psi)
&=8\pi\,\delta\pi_{ij}^{\rm(E)}\big|_S.
\label{eq:ScalarAniEinSide}
\end{align}

Equations~\eqref{eq:ScalarHamEinSide}--\eqref{eq:ScalarAniEinSide} are the scalar energy constraint, momentum constraint, trace-$ij$ equation, and scalar anisotropy equation, respectively, now written in direct correspondence with the effective density, heat-flux, pressure, and anisotropic-stress channels.

Inserting the effective split gives
\begin{eqnarray}
3H(\dot\psi+HA)-\frac{\nabla^2}{a^2}\psi
&=&-\frac{4\pi}{\phib}\delta\rho_m+\frac{4\pi\varphi}{\phib^2}\bar\rho_m-4\pi\delta\rho_\phi,\nonumber \\
\label{eq:ScalarHamWithPhi}\\
\partial_i(\dot\psi+HA-\dot Hv_\phi)
&=&-\frac{4\pi}{\phib}\delta q_i^{(m)}\big|_S-4\pi\delta q_i^{(\phi)},
\label{eq:ScalarMomWithPhi}\\
\tilde{H}+\tilde{\psi}-\frac{1}{3a^2}\nabla^2(A-\psi)
&=&-\frac{4\pi}{\phib}\delta P_m+\frac{4\pi\varphi}{\phib^2}\bar P_m-4\pi\delta P_\phi,\nonumber \\
\label{eq:ScalarPressWithPhi}\\
-\Dij(A-\psi)
&=&\frac{8\pi}{\phib}\delta\pi_{ij}^{(m)}\big|_S+8\pi\delta\pi_{ij}^{(\phi)}\big|_S \,,
\label{eq:ScalarAniWithPhi}
\end{eqnarray}
where $\tilde{\psi}\equiv \psi(2\dot{H}+3H^2)$ for simplification.

We now focus on the two genuinely imperfect channels, namely the projected heat flux and the scalar anisotropic stress. Inserting the explicit scalar--tensor contributions from Sec.~\ref{sec:thermalVars} into Eqs.~\eqref{eq:ScalarMomWithPhi} and \eqref{eq:ScalarAniWithPhi} yields
\begin{align}
\partial_i(\dot\psi+HA-\dot H\,v_\phi)
&=-\frac{4\pi}{\phib}\delta q_i^{(m)}\big|_S
+4\pi\,\bar\kappa\bar T\,\partial_i(A+\dot v_\phi),
\label{eq:ScalarMomWithPhiFullyExplicit}\\
-\Dij(A-\psi)
&=\frac{8\pi}{\phib}\delta\pi_{ij}^{(m)}\big|_S+\frac{1}{\phib}\Dij\varphi.
\label{eq:ScalarAniWithPhiFullyExplicit}
\end{align}
where we used the acceleration-driven form of the scalar--tensor heat flux~\eqref{eq:dqCompact} and the perturbed 4-acceleration given by Eq.~\eqref{eq:deltaaiStart} together with the scalar anisotropic-stress relation~\eqref{eq:piScalarFinal}.

These equations show that the scalar sector of Jordan-frame scalar--tensor gravity couples to the metric not only through effective density and pressure perturbations, but also through two intrinsically imperfect-fluid channels. First, the projected momentum equation is sourced by an effective heat flux whose scalar--tensor contribution is acceleration-driven. Second, the scalar anisotropy equation is sourced by an intrinsic scalar anisotropic stress proportional to $\Dij\varphi$. As a consequence, even in the absence of ordinary-matter anisotropic stress, the scalar perturbation $\varphi$ generically produces a nonvanishing gravitational slip $A-\psi$. In this precise sense, the perturbed modified-gravity sector behaves as an effective imperfect medium for scalar perturbations.

A particularly transparent limit is the ordinary-matter vacuum case, for which
\begin{equation}
\delta q_i^{(m)}\big|_S=0,
\qquad
\delta\pi_{ij}^{(m)}\big|_S=0.
\end{equation}
Equations~\eqref{eq:ScalarMomWithPhiFullyExplicit} and \eqref{eq:ScalarAniWithPhiFullyExplicit} then reduce to
\begin{equation}
\partial_i(\dot\psi+HA-\dot H\,v_\phi)
=
4\pi\,\bar\kappa\bar T\,\partial_i(A+\dot v_\phi),
\end{equation}
\begin{equation}
-\Dij(A-\psi)=\frac{1}{\bar\phi}\Dij\varphi.
\end{equation}

Therefore, even in the absence of ordinary matter, the scalar degree of freedom acts as an intrinsic source of both a heat-flux channel and an anisotropic-stress channel in the perturbed Einstein-like system. In particular, the scalar perturbation $\varphi$ generically generates a nonzero gravitational slip already in vacuum.

By contrast, at a local constant-background vacuum point with
\begin{equation}
\dot{\bar\phi}=0,
\qquad
\ddot{\bar\phi}=0,
\end{equation}
the scalar-gradient frame ceases to be defined and the heat-flux interpretation based on $v_\phi=-\varphi/\dot{\bar\phi}$ is no longer available. In that limit the scalar mode is more naturally described by the Klein--Gordon-type equation.

\subsection{Tensor gravitational waves and the TT thermal channel}\label{sec:tensorwave}

We now turn to the tensor gravitational-wave sector and isolate the transverse-traceless perturbations that propagate independently of the scalar modes at linear order. For pure tensors,
\begin{equation}
A=\psi=\varphi=0,
\qquad
\chi_{ij}\neq 0,
\qquad
\partial^i\chi_{ij}=0,
\qquad
\chi^i{}_i=0,
\end{equation}
leading to the mixed-index TT Einstein equation
\begin{equation}
\delta G^i{}_j\big|_{\TT}=8\pi\,\delta\Tmath^i{}_j\big|_{\TT},
\label{eq:TensorEinSideTmathfrak}
\end{equation}
with
\begin{equation}
\delta\Tmath^i{}_j\big|_{\TT}=\frac{1}{\phib}\delta T^{(m)i}{}_j\big|_{\TT}+\delta T^{(\phi)i}{}_j\big|_{\TT}.
\label{eq:TensorSourceSplitTmathfrak}
\end{equation}

The effective TT source computed in Sec.~\ref{sec:thermalVars} obeys
\begin{equation}
8\pi\,\delta T^{(\phi)i}{}_j\big|_{\TT}=8\pi\,\delta\pi^{(\phi)i}{}_j\big|_{\TT}=-\frac12\frac{\dot\phib}{\phib}\dot\chi^i{}_j=4\pi\bar{\kappa}\bar{T}\dot\chi^i{}_j.
\label{eq:TensorPhiMixedTT}
\end{equation}
and using the standard TT Einstein operator then yields
\begin{equation}
\ddot\chi^i{}_j+\left(3H-8\pi\bar{\kappa}\bar{T}\right)\dot\chi^i{}_j-\frac{\nabla^2}{a^2}\chi^i{}_j=-\frac{16\pi}{\phib}\delta T^{(m)i}{}_j\big|_{\TT}.
\label{eq:TensorWithMatterTTFinal}
\end{equation}

To cast the tensor equation in a sourced form with a positive right-hand side, we define
\begin{equation}
\Pi^{(m)i}{}_j \equiv -\,\delta T^{(m)i}{}_j\big|_{\TT},
\end{equation}
where this sign choice is purely conventional and should not be confused with the imperfect-fluid anisotropic-stress tensor $\pi_{ij}$ used elsewhere in the paper. In many cosmology references the TT source is defined without this minus sign, in which case Eq.~\eqref{eq:TensorGWFinal} would be written with a minus sign on the right-hand side. Hence, with $\Pi^{(m)i}{}_j\equiv -\delta T^{(m)i}{}_j|_{\TT}$ one can write
\begin{equation}
\ddot\chi^i{}_j+\left(3H-8\pi\bar{\kappa}\bar{T}\right)\dot\chi^i{}_j-\frac{\nabla^2}{a^2}\chi^i{}_j=\frac{16\pi}{\phib}\Pi^{(m)i}{}_j.
\label{eq:TensorGWFinal}
\end{equation}

While the Jordan-frame tensor propagation equation is standard in scalar--tensor cosmological perturbation theory, the present analysis provides an exact perturbative thermal-channel reconstruction of that equation. In particular, the additional friction term can be identified with the TT anisotropic-stress contribution of the effective $\phi$-fluid, thereby linking the familiar gravitational-wave propagation law to the recent Eckart-type thermodynamic interpretation of scalar--tensor gravity. Written in terms of the effective thermal quantity $\bar\kappa\bar T$, the tensor equation makes explicit that the scalar--tensor correction to gravitational-wave propagation is itself an exact constitutive channel of the effective-fluid description. Indeed, the damping coefficient
\begin{equation}
\Gamma_{\rm gw}=3H-8\pi\bar\kappa\bar T,
\end{equation}
contains, in addition to the standard Hubble dilution term, a purely scalar--tensor contribution that coincides precisely with the TT anisotropic-stress channel of the effective $\phi$ sector. In this way, the same effective thermal variable that governs the scalar heat-flux channel also controls the modified damping of tensor modes.

This rewriting also clarifies several physically relevant limits. In matter vacuum, the thermal term survives and therefore isolates the pure modified-gravity contribution to gravitational-wave propagation, whereas for $\dot{\bar\phi}=0$ one has $\bar\kappa\bar T=0$ and the standard GR damping law is recovered. In the source-free subhorizon regime, the tensor amplitude scales as $\chi_{ij}\propto (a\sqrt{\bar\phi})^{-1}$ rather than $\chi_{ij}\propto a^{-1}$, so that $\bar\kappa\bar T$ directly measures the departure from the usual GR redshifting behaviour. Moreover, on the future-directed branch, for which $8\pi\bar\kappa\bar T=-\dot{\bar\phi}/\bar\phi>0$, the scalar--tensor sector reduces the usual Hubble damping. If the thermal contribution becomes comparable to the cosmological one, $8\pi\bar\kappa\bar T\simeq 3H$, the net damping can be strongly suppressed; if $8\pi\bar\kappa\bar T>3H$, the tensor equation enters an anti-friction regime.

The entropy analysis also clarifies the scope of this thermal interpretation. At linear order, the tensor sector already displays an exact transport-like channel through the identification of the modified damping term with the TT anisotropic stress of the effective $\phi$ fluid. However, because entropy production is quadratic in the dissipative quantities and vanishes identically on FLRW at first order, the linear tensor analysis does not by itself establish genuinely irreversible thermodynamic behaviour. Rather, it isolates the precise channel through which such behaviour could manifest. In this sense, the present TT result may be viewed as evidence that scalar--tensor gravitational waves admit a nontrivial effective thermal description at the constitutive level, while a decisive assessment of deeper thermal properties must await a second-order analysis.

\section{Scalar-wave equation and thermal-channel interpretation}\label{sec:scalarwave}

Perturbing the scalar-field equation \eqref{eq:scalarFE} around a spatially flat FLRW background gives
\begin{align}
	&(2\omegab+3)\,\delta(\Box\phi)+2\omegapb\,\varphi\,\bar\Box\bar\phi
	\nonumber\\
	&\qquad
	=
	8\pi\,\delta T^{(m)}
	-\delta\!\left[\omega_{,\phi}(\nabla\phi)^2\right]
	+\delta\!\left[2\phi V_{,\phi}-4V\right].
	\label{eq:scalarPertStart_expanded}
\end{align}

Using the background relation
\begin{equation}
	\bar\Box\bar\phi=-\ddot\phib-3H\dot\phib,
\end{equation}
together with the standard linearized expressions
\begin{equation}
\delta(\Box\phi)=
-\ddot\varphi-3H\dot\varphi+\frac{\nabla^2}{a^2}\varphi
+\dot\phib(\dot A+3\dot\psi)
+2A(\ddot\phib+3H\dot\phib),
\label{eq:deltaBoxPhi_scalarwave}
\end{equation}
and
\begin{equation}
\delta\!\left[\omega_{,\phi}(\nabla\phi)^2\right]
=
-\omegappb\dot\phib^2\varphi
+\omegapb\left(2A\dot\phib^2-2\dot\phib\dot\varphi\right),
\label{eq:deltaomegaGradPhi2}
\end{equation}
as well as
\begin{equation}
\delta\!\left[2\phi V_{,\phi}-4V\right]
=
2(\phib\Vppb-\Vpb)\varphi,
\label{eq:deltaPotentialCombination}
\end{equation}
one obtains
\begin{widetext}
\begin{eqnarray}
(2\omegab+3)\Big(
-\ddot\varphi-3H\dot\varphi+\frac{\nabla^2}{a^2}\varphi
+\dot\phib(\dot A+3\dot\psi)
+2A(\ddot\phib+3H\dot\phib)
\Big)
&=&
8\pi\,\delta T^{(m)}
+2\omegapb(\ddot\phib+3H\dot\phib)\varphi
+\omegappb\dot\phib^2\varphi
	\nonumber\\
&& \hspace{-3cm} -2\omegapb A\dot\phib^2
+2\omegapb\dot\phib\dot\varphi
+2(\phib\Vppb-\Vpb)\varphi ,
\label{eq:scalarPertFull_section}
\end{eqnarray}
which can be rewritten equivalently as
\begin{align}
\ddot\varphi+3H\dot\varphi-\frac{\nabla^2}{a^2}\varphi
&= \dot\phib(\dot A+3\dot\psi)+2A(\ddot\phib+3H\dot\phib)
-\frac{8\pi}{2\omegab+3}\,\delta T^{(m)}
-\frac{2\omegapb}{2\omegab+3}(\ddot\phib+3H\dot\phib)\varphi
\nonumber\\
&\quad
-\frac{\omegappb\dot\phib^2}{2\omegab+3}\,\varphi
+\frac{2\omegapb\dot\phib^2}{2\omegab+3}\,A
-\frac{2\omegapb\dot\phib}{2\omegab+3}\,\dot\varphi
-\frac{2(\phib\Vppb-\Vpb)}{2\omegab+3}\,\varphi .
\label{eq:scalarWaveGeneral_section}
\end{align}

\end{widetext}

Equation~\eqref{eq:scalarWaveGeneral_section} shows that the scalar mode is not governed by
$\varphi$ alone: it is coupled to the metric potentials $A$ and $\psi$, and therefore to the
full perturbed Einstein-like system. More precisely, $A$ and $\psi$ are fixed by the effective
density, pressure, heat-flux, and anisotropic-stress channels through
Eqs.~\eqref{eq:ScalarHamEinSide}--\eqref{eq:ScalarAniEinSide}. In this sense, although the
scalar-wave equation is not written directly in terms of the thermal variables, it is controlled
indirectly and exactly by the same effective channels that organize the perturbed Einstein-like
equations.

This channel structure is particularly transparent when combined with the constitutive results of
Secs.~\ref{sec:thermalVars} and \ref{sec:EckartConnection}. In the scalar sector, the effective
heat flux takes the acceleration-driven form
\begin{equation}
\delta q_i^{(\phi)}
=
-\bar\kappa\bar T\,\delta a_i,
\qquad
\delta a_i=\partial_i(A+\dot v_\phi),
\label{eq:qphiThermalScalarWaveSec}
\end{equation}
while the scalar anisotropic stress is
\begin{equation}
8\pi\,\delta\pi_{ij}^{(\phi)}\big|_S
=
\frac{1}{\phib}\Dij\varphi
=
-\,\frac{\dot\phib}{\phib}\,\delta\sigma_{ij}\big|_S.
\label{eq:piphiThermalScalarWaveSec}
\end{equation}
Accordingly, the heat-flux channel enters the scalar-wave problem through the momentum constraint,
the anisotropic-stress channel through the gravitational slip $A-\psi$, and the density and
pressure channels through the Hamiltonian and trace-$ij$ equations. Unlike the tensor sector,
where the TT anisotropic stress isolates a single propagation channel, the scalar sector is
therefore dressed by all four effective thermal channels of the Einstein-like decomposition.

A particularly transparent limit is the matter-vacuum case,
\begin{equation}
\delta\rho_m=\delta P_m=\delta q_i^{(m)}=\delta\pi_{ij}^{(m)}=0.
\label{eq:matterVacuumScalar}
\end{equation}
Then the scalar perturbation is governed entirely by the effective $\phi$-sector channels. In
this case the modified-gravity sector acts as a self-generated effective medium for scalar
perturbations: the momentum transport is supplied by the acceleration-driven scalar heat flux, the
gravitational slip is supplied by the scalar anisotropic stress, and the Hamiltonian and trace-$ij$
equations are sourced solely by the scalar effective density and pressure.

A different special case occurs at a local constant-background vacuum point,
\begin{equation}
\dot\phib=0,
\qquad
\ddot\phib=0,
\qquad
T^{(m)}=0,
\label{eq:localVacuumPoint}
\end{equation}
for which the background condition is $2\phib\Vpb-4\Vb=0$. In this limit the scalar-gradient
congruence is no longer defined, so the heat-flux interpretation based on
$v_\phi=-\varphi/\dot\phib$ ceases to apply. The scalar perturbation nevertheless survives as an
ordinary Klein--Gordon-type mode,
\begin{equation}
(2\omegab+3)\Box\varphi-2(\phib\Vppb-\Vpb)\varphi=0,
\end{equation}
or equivalently
\begin{equation}
\left(\Box-m_{\rm eff}^2\right)\varphi=0,
\qquad
m_{\rm eff}^2=\frac{2(\phib\Vppb-\Vpb)}{2\omegab+3}.
\label{eq:scalarKleinGordon_section}
\end{equation}

The physical picture is therefore the following. Away from the constant-background vacuum point,
the scalar degree of freedom does not propagate as an autonomous Klein--Gordon field on a fixed
FLRW background. Rather, it propagates as a mode coupled to the metric sector, and the latter is
itself organized by the four effective thermal channels of the Einstein-like decomposition. In
this precise sense, scalar-wave propagation in Jordan-frame scalar--tensor gravity is governed by
the full constitutive structure of the effective imperfect-fluid description.

\section{Conclusion}\label{sec:Discussion}

In this work we developed a first-order perturbative analysis of Jordan-frame scalar--tensor gravity on a spatially flat FLRW background, organized around the Einstein-like effective-fluid decomposition of the scalar sector. The central result is that the perturbative effective density, pressure, heat flux, and anisotropic stress admit an exact Eckart-type constitutive identification in the scalar-gradient frame, and that this constitutive structure is realized explicitly in the complete linearized field equations. The novelty does not lie in the mere appearance of fluid-like slots in a projected Einstein-like system, since that is a generic consequence of the $1+3$ decomposition. Rather, the nontrivial statement established here is that these channels possess a precise constitutive meaning and govern the full first-order scalar and tensor perturbation dynamics.

In the scalar sector, the linearized Einstein-like equations admit a complete channel-by-channel organization: the Hamiltonian, momentum, trace, and scalar traceless equations are governed, respectively, by the effective density, heat-flux, pressure, and anisotropic-stress channels. In particular, the momentum equation contains an acceleration-driven effective heat flux, while the anisotropy equation contains an intrinsic scalar anisotropic stress proportional to $\Dij\varphi$. As a result, the scalar degree of freedom behaves as an effective imperfect medium already at first order and generically induces nonvanishing gravitational slip even in the absence of ordinary-matter anisotropic stress.

The tensor sector yields the sharpest constitutive result. The standard Jordan-frame gravitational-wave propagation law is recovered, but its extra damping term can be identified exactly with the transverse-traceless anisotropic-stress channel of the effective scalar sector. Written in terms of the effective quantity $\bar\kappa\bar T$, the tensor equation shows that the same gravity-side thermal variable that controls the scalar heat-flux channel also controls the modified damping of tensor modes. In this sense, the tensor propagation law is not merely compatible with the effective thermodynamic picture, but provides its most direct realization at linear order.

The perturbed scalar-field equation provides the complementary part of the picture. Away from special limits, the scalar mode does not propagate as an autonomous Klein--Gordon perturbation on a fixed FLRW background. Rather, it remains coupled to the metric sector, and hence indirectly to the same effective density, pressure, heat-flux, and anisotropic-stress channels that organize the Einstein-like equations. Scalar-wave propagation is therefore governed by the full imperfect-fluid structure of the effective scalar sector, whereas the tensor sector isolates a single channel in a particularly transparent way.

A further important result is the evolution equation for $\delta(\kappa T)$. On FLRW, matching the perturbed heat flux fixes only the background coefficient $\overline{\kappa T}$; it does not determine $\delta(\kappa T)$ itself. The latter obeys an independent dynamical equation and carries additional perturbative information sourced by expansion, lapse perturbations, and matter/potential terms. This shows that the thermal reconstruction is not merely a relabeling of the momentum equation, but includes a distinct evolution law for the gravity-fixed thermal scalar.

The analysis also clarifies the scope of the thermodynamic interpretation. The theory fixes the invariant product $\kappa T$, not $\kappa$ and $T$ separately, so any separate split into conductivity and temperature is constitutive rather than fundamental. Likewise, although the effective anisotropic stress obeys an exact Eckart/Landau--Lifshitz-type shear law and the effective heat flux takes an acceleration-driven Eckart form under the appropriate projected-gradient condition, these results establish a constitutive structure rather than microscopic thermality. The entropy discussion reinforces this point: because entropy production is quadratic in the dissipative quantities, its linear perturbation vanishes identically on FLRW, and the first nontrivial contribution arises only at second order. Thus the linear theory already reveals the full transport-like structure of the effective scalar sector, but not yet a genuinely irreversible entropy balance.

We therefore view the present work as establishing a precise perturbative framework in which stronger questions can now be posed. The exact constitutive reconstruction obtained here shows that the thermal viewpoint captures genuine structural features of scalar--tensor perturbation theory. In particular, it leaves open the possibility that gravitational waves in scalar--tensor gravity admit a deeper thermodynamic characterization, perhaps even an intrinsic one, although the present analysis establishes this only at the level of an effective constitutive description. The natural next step is to extend the analysis beyond first order, where entropy production, nonlinear channel coupling, and genuinely nontrivial transport effects first appear. That is the appropriate setting in which to determine whether the constitutive structure identified here remains purely organizational or points toward a deeper thermodynamic characterization of scalar--tensor gravitational perturbations.

\appendix

\section{Detailed and essential derivations for the perturbation calculations}\label{app:geom}

\subsection{Metric and connections}

Starting from the metric~\eqref{eq:metricPert} we write $g^{ab}=\bar g^{ab}+\delta g^{ab}$ and impose $g^{ac}g_{cb}=\delta^a{}_b$. To first order,
\begin{equation}
\delta g^{ac}\,\bar g_{cb}+\bar g^{ac}\,\delta g_{cb}=0.
\end{equation}
For $(00)$,
\begin{equation}
\delta g^{00}\,(-1)+(-1)(-2A)=0 \quad\Rightarrow\quad \delta g^{00}=2A.
\end{equation}
For $(ij)$,
\begin{equation}
\delta g^{ij}a^2\delta_{jk}+a^{-2}\delta^{ij}a^2(-2\psi\delta_{jk}+\chi_{jk})=0,
\end{equation}
so
\begin{equation}
\delta g^{ij}=a^{-2}(2\psi\delta^{ij}-\chi^{ij}).
\end{equation}

For the Christoffel symbols,
\begin{equation}
\Gamma^a{}_{bc}=\frac12 g^{ad}(\partial_b g_{cd}+\partial_c g_{bd}-\partial_d g_{bc}).
\end{equation}
Hence
\begin{equation}
\delta\Gamma^0{}_{00}=\frac12\bar g^{00}(2\partial_0\delta g_{00}-\partial_0\delta g_{00})=\dot A,
\end{equation}
and
\begin{equation}
\delta\Gamma^0{}_{0i}=\frac12\bar g^{00}\partial_i\delta g_{00}=\partial_i A.
\end{equation}
For the spatial component,
\begin{align}
\Gamma^0{}_{ij}&=-\frac12 g^{00}\partial_0 g_{ij}
\qquad (g_{0i}=0),
\end{align}
so the first-order perturbation is
\begin{align}
\delta\Gamma^0{}_{ij}
&=-\frac12\delta g^{00}\partial_0\bar g_{ij}-\frac12\bar g^{00}\partial_0\delta g_{ij}\nonumber\\
&=-A(2a^2H\delta_{ij})+\frac12\partial_0\left[a^2(-2\psi\delta_{ij}+\chi_{ij})\right].
\end{align}

\subsection{Derivation of the $3+1$ perturbed quantities}\label{app:perturbed31}

Starting with
\begin{equation}
\delta a_i=
\bar u^b\,\delta(\nabla_b u_i)
+\delta u^b\,\bar\nabla_b \bar u_i.
\label{eq:deltaai}
\end{equation}
one gets
\begin{equation}
\bar u^b\,\delta(\nabla_b u_i)
=
\delta(\nabla_0 u_i)
=
\partial_0 \delta u_i
-\bar\Gamma^j{}_{0i}\delta u_j
-\delta\Gamma^0{}_{0i}\bar u_0,
\end{equation}
since $\bar\Gamma^0{}_{0i}=0$, $\bar u_0=-1$. Therefore
\begin{equation}
\bar u^b\,\delta(\nabla_b u_i)
=
\partial_i\dot v_\phi - H\partial_i v_\phi + \partial_i A.
\label{eq:deltaaiPart1}
\end{equation}
For the second term in \eqref{eq:deltaaiStart},
\begin{equation}
\bar\nabla_j\bar u_i = -\bar\Gamma^0{}_{ji}\bar u_0 = Ha^2\delta_{ji},
\qquad
\delta u^j=\bar g^{jk}\delta u_k = a^{-2}\partial^j v_\phi,
\end{equation}
hence
\begin{equation}
\delta u^b\,\bar\nabla_b\bar u_i
=
\delta u^j\,\bar\nabla_j\bar u_i
=
H\partial_i v_\phi.
\label{eq:deltaaiPart2}
\end{equation}
Adding \eqref{eq:deltaaiPart1} and \eqref{eq:deltaaiPart2}:
\begin{equation}
\delta a_i = \partial_i\left(A+\dot v_\phi\right).
\label{eq:deltaaiFinal}
\end{equation}

\subsection{Derivation of \texorpdfstring{$\delta[(\nabla\phi)^2]$}{delta(grad phi)^2} and \texorpdfstring{$\delta(\Box\phi)$}{delta Box phi}}
\label{app:operators}

First,
\begin{equation}
(\nabla\phi)^2=g^{ab}\partial_a\phi\partial_b\phi.
\end{equation}
Linearizing,
\begin{align}
\delta[(\nabla\phi)^2]
&=\delta g^{ab}\partial_a\bar\phi\partial_b\bar\phi
+2\bar g^{ab}\partial_a\bar\phi\partial_b\varphi.
\end{align}
Since $\partial_i\bar\phi=0$ and $\bar g^{00}=-1$,
\begin{equation}
\delta[(\nabla\phi)^2]
=\delta g^{00}\dot\phib^2+2\bar g^{00}\dot\phib\dot\varphi
=2A\dot\phib^2-2\dot\phib\dot\varphi.
\end{equation}

For the d'Alembertian,
\begin{equation}
\Box\phi=g^{ab}\nabla_a\nabla_b\phi,
\end{equation}
so its first-order perturbation is
\begin{equation}
\delta(\Box\phi)
=
\delta g^{ab}\,\bar\nabla_a\bar\nabla_b\bar\phi
+\bar g^{ab}\,\delta(\nabla_a\nabla_b\phi).
\label{eq:deltaBox_split}
\end{equation}

Because $\bar\phi=\bar\phi(t)$, the nonzero background second derivatives are
\begin{equation}
\bar\nabla_0\bar\nabla_0\bar\phi=\ddot\phib,
\qquad
\bar\nabla_i\bar\nabla_j\bar\phi
=-\bar\Gamma^0{}_{ij}\dot\phib
=-a^2H\delta_{ij}\dot\phib.
\end{equation}
Using
\begin{equation}
\delta g^{00}=2A,
\qquad
\delta g^{ij}=a^{-2}(2\psi\,\delta^{ij}-\chi^{ij}),
\end{equation}
one gets
\begin{align}
\delta g^{ab}\,\bar\nabla_a\bar\nabla_b\bar\phi
&=
\delta g^{00}\ddot\phib+\delta g^{ij}\bar\nabla_i\bar\nabla_j\bar\phi
\nonumber\\
&=
2A\ddot\phib
+a^{-2}(2\psi\,\delta^{ij}-\chi^{ij})(-a^2H\delta_{ij}\dot\phib)
\nonumber\\
&=
2A\ddot\phib-6H\psi\,\dot\phib,
\label{eq:deltaBox_metricpart}
\end{align}
where the TT part drops out because $\chi^i{}_i=0$.

Next, for a scalar field,
\begin{equation}
\delta(\nabla_a\nabla_b\phi)
=
\bar\nabla_a\bar\nabla_b\varphi
-\delta\Gamma^c{}_{ab}\bar\nabla_c\bar\phi.
\label{eq:delta_second_derivative_scalar}
\end{equation}
Since only $\bar\nabla_0\bar\phi=\dot\phib$ is nonzero, this becomes
\begin{align}
\delta(\nabla_0\nabla_0\phi)
&=
\ddot\varphi-\delta\Gamma^0{}_{00}\dot\phib,
\label{eq:delta_nabla00phi}
\\
\delta(\nabla_i\nabla_j\phi)
&=
\bar\nabla_i\bar\nabla_j\varphi-\delta\Gamma^0{}_{ij}\dot\phib.
\label{eq:delta_nablaijphi_step}
\end{align}
Now,
\begin{equation}
\bar\nabla_i\bar\nabla_j\varphi
=
\partial_i\partial_j\varphi-\bar\Gamma^0{}_{ij}\dot\varphi,
\end{equation}
because in Cartesian spatial coordinates for flat FLRW one has $\bar\Gamma^k{}_{ij}=0$ and
\begin{equation}
\bar\Gamma^0{}_{ij}=a^2H\delta_{ij}.
\end{equation}
Therefore,
\begin{equation}
\delta(\nabla_i\nabla_j\phi)
=
\partial_i\partial_j\varphi
-\delta\Gamma^0{}_{ij}\dot\phib
-a^2H\delta_{ij}\dot\varphi.
\label{eq:delta_nablaijphi}
\end{equation}

We also need the perturbed Christoffel symbols. For the time-time component,
\begin{equation}
\delta\Gamma^0{}_{00}=\dot A.
\label{eq:deltaGamma000_app}
\end{equation}
For the spatial component,
\begin{align}
\delta\Gamma^0{}_{ij}
&=
-\frac12\delta g^{00}\partial_0\bar g_{ij}
-\frac12\bar g^{00}\partial_0\delta g_{ij}
\nonumber\\
&=
-2a^2HA\,\delta_{ij}
+\frac12\partial_0\!\left[a^2(-2\psi\,\delta_{ij}+\chi_{ij})\right]
\nonumber\\
&=
-a^2\left(2HA+2H\psi+\dot\psi\right)\delta_{ij}
+a^2H\chi_{ij}
+\frac{a^2}{2}\dot\chi_{ij}.
\label{eq:deltaGamma0ij_app}
\end{align}
Hence
\begin{equation}
a^{-2}\delta^{ij}\delta\Gamma^0{}_{ij}
=
-6HA-6H\psi-3\dot\psi.
\label{eq:trace_deltaGamma0ij_app}
\end{equation}

We can now contract Eq.~\eqref{eq:delta_second_derivative_scalar}. Using
$\bar g^{00}=-1$ and $\bar g^{ij}=a^{-2}\delta^{ij}$,
\begin{align}
\bar g^{ab}\,\delta(\nabla_a\nabla_b\phi)
&=
-\delta(\nabla_0\nabla_0\phi)
+a^{-2}\delta^{ij}\delta(\nabla_i\nabla_j\phi)
\nonumber\\
&=
-\left(\ddot\varphi-\dot A\,\dot\phib\right) \nonumber\\
&+a^{-2}\delta^{ij} \left( \partial_i\partial_j\varphi -\delta\Gamma^0{}_{ij}\dot\phib
-a^2H\delta_{ij}\dot\varphi
\right)
\nonumber\\
&=
-\ddot\varphi+\dot A\,\dot\phib
+\frac{\nabla^2}{a^2}\varphi
-\dot\phib\,a^{-2}\delta^{ij}\delta\Gamma^0{}_{ij}
-3H\dot\varphi
\nonumber\\
&= -\ddot\varphi-3H\dot\varphi+\frac{\nabla^2}{a^2}\varphi \nonumber \\
&+\dot\phib\left(\dot A+6HA+6H\psi+3\dot\psi\right).
\label{eq:deltaBox_connectionpart}
\end{align}

Finally, combining Eqs.~\eqref{eq:deltaBox_metricpart} and \eqref{eq:deltaBox_connectionpart}
in Eq.~\eqref{eq:deltaBox_split}, the $6H\psi\dot\phib$ terms cancel and one obtains
\begin{equation}\label{eq:deltaBoxPhi}
\delta(\Box\phi)
= -\ddot\varphi-3H\dot\varphi+\frac{\nabla^2}{a^2}\varphi +\dot\phib\left(\dot A+3\dot\psi\right) +2A\left(\ddot\phib+3H\dot\phib\right).
\end{equation}

\subsection{Detailed derivation of the effective thermal variables}\label{app:thermalDerivations}

We split
\begin{equation}
8\pi T^{(\phi)}_{ab}=X_{ab}+Y_{ab}+Z_{ab},
\end{equation}
with
\begin{align}
X_{ab}&=\frac{\omega(\phi)}{\phi^2}\left(\nabla_a\phi\nabla_b\phi-\frac12 g_{ab}(\nabla\phi)^2\right),\\
Y_{ab}&=\frac{1}{\phi}\left(\nabla_a\nabla_b\phi-g_{ab}\Box\phi\right),\\
Z_{ab}&=-\frac{V}{\phi}g_{ab}.
\end{align}

\subsubsection{The \texorpdfstring{$(0,0)$}{(0,0)} component}

Define $F(\phi)=\omega(\phi)/\phi^2$. Then
\begin{equation}
\delta F=\left(\frac{\omegapb}{\phib^2}-\frac{2\omegab}{\phib^3}\right)\varphi.
\end{equation}
Also
\begin{equation}
B_{00}\equiv \nabla_0\phi\nabla_0\phi-\frac12 g_{00}(\nabla\phi)^2,
\qquad
\bar B_{00}=\frac12\dot\phib^2.
\end{equation}
Now
\begin{align}
\delta B_{00}
&=2\dot\phib\dot\varphi-\frac12\delta g_{00}(\nabla\bar\phi)^2-\frac12\bar g_{00}\delta[(\nabla\phi)^2]\nonumber\\
&=2\dot\phib\dot\varphi-\frac12(-2A)(-\dot\phib^2)-\frac12(-1)(2A\dot\phib^2-2\dot\phib\dot\varphi)\nonumber\\
&=\dot\phib\dot\varphi.
\end{align}
Hence
\begin{align}
\delta X_{00}&=\delta F\bar B_{00}+\bar F\,\delta B_{00} \nonumber \\
&=\left(\frac{\omegapb}{2\phib^2}-\frac{\omegab}{\phib^3}\right)\dot\phib^2\varphi+\frac{\omegab}{\phib^2}\dot\phib\dot\varphi.
\end{align}

Next define
\begin{equation}
C_{00}\equiv \nabla_0\nabla_0\phi-g_{00}\Box\phi,
\qquad
\bar C_{00}=-3H\dot\phib.
\end{equation}
Using Eq.~\eqref{eq:deltaBoxPhi} one finds
\begin{equation}
\delta C_{00}=\frac{\nabla^2}{a^2}\varphi-3H\dot\varphi+3\dot\phib\dot\psi.
\end{equation}
Therefore
\begin{align}
\delta Y_{00}&=-\frac{\varphi}{\phib^2}\bar C_{00}+\frac{1}{\phib}\delta C_{00} \nonumber\\
&=\frac{3H\dot\phib}{\phib^2}\varphi+\frac{1}{\phib}\left(\frac{\nabla^2}{a^2}\varphi-3H\dot\varphi+3\dot\phib\dot\psi\right).
\end{align}
Finally,
\begin{equation}
\delta Z_{00}=\left(\frac{\Vpb}{\phib}-\frac{\Vb}{\phib^2}\right)\varphi+2A\frac{\Vb}{\phib}.
\end{equation}
Summing and using Eq.~\eqref{eq:deltaRhoProjection} produces Eq.~\eqref{eq:deltaRhoFinal}.

\subsubsection{The \texorpdfstring{$(0,i)$}{(0,i)} component}

Since $g_{0i}=0$,
\begin{equation}
8\pi\,\delta T^{(\phi)}_{0i}=\delta\left[\frac{\omega}{\phi^2}\nabla_0\phi\nabla_i\phi\right]+\delta\left[\frac{1}{\phi}\nabla_0\nabla_i\phi\right].
\end{equation}
Because $\nabla_i\bar\phi=0$,
\begin{equation}
\delta\left[\frac{\omega}{\phi^2}\nabla_0\phi\nabla_i\phi\right]=\frac{\omegab}{\phib^2}\dot\phib\,\partial_i\varphi.
\end{equation}
Also,
\begin{align}
\delta(\nabla_0\nabla_i\phi)
&=\partial_i\dot\varphi-\delta\Gamma^0{}_{0i}\dot\phib-\bar\Gamma^j{}_{0i}\partial_j\varphi\nonumber\\
&=\partial_i\dot\varphi-\dot\phib\,\partial_iA-H\partial_i\varphi.
\end{align}

\subsubsection{The spatial \texorpdfstring{$(i,j)$}{(i,j)} components}

The full scalar part of $\delta T^{(\phi)}_{ij}$ can be written as
\begin{equation}
\delta T^{(\phi)}_{ij}\big|_S=a^2\left(\delta P_\phi-2\Pb\psi\right)\delta_{ij}+\frac{1}{8\pi\phib}\Dij\varphi.
\label{eq:scalarTijFinalAppendix}
\end{equation}

The key point is that the only scalar tracefree contribution comes from the $\partial_i\partial_j\varphi$ piece inside $\delta(\nabla_i\nabla_j\phi)$, while all remaining scalar terms are proportional to $\delta_{ij}$. Likewise, in the pure TT sector,
\begin{equation}
8\pi\,\delta T^{(\phi)}_{ij}\big|_{\TT}=8\pi\Pb\,a^2\chi_{ij}-\frac{a^2}{2}\frac{\dot\phib}{\phib}\dot\chi_{ij},
\end{equation}
so Eq.~\eqref{eq:deltaPiProjection} gives Eq.~\eqref{eq:piTTMain} immediately.

\subsection{Derivation of the shear-anisotropic-stress relation}\label{app:shear}

Starting from Eq.~\eqref{eq:shear_def}, the linear spatial shear is
\begin{equation}
\delta\sigma_{ij}=\left[\delta(\nabla_{(i}u_{j)})-H\delta g_{ij}\right]^{\TF}.
\end{equation}
Now
\begin{equation}
\delta(\nabla_i u_j)=\partial_i\delta u_j-\delta\Gamma^c{}_{ij}\bar u_c-\bar\Gamma^c{}_{ij}\delta u_c.
\end{equation}
Using $\bar u_0=-1$, $\bar u_k=0$, and $\bar\Gamma^0{}_{ij}=a^2H\delta_{ij}$ gives
\begin{equation}
\delta(\nabla_i u_j)=\partial_i\delta u_j+\delta\Gamma^0{}_{ij}-a^2H\delta_{ij}\delta u_0.
\end{equation}
The last term is pure trace and disappears after TF projection. In the scalar sector, $\delta u_i=\partial_i v_\phi$ and both $\delta\Gamma^0{}_{ij}\big|_S$ and $\delta g_{ij}\big|_S$ are pure trace, leading to
\begin{equation}
\delta\sigma_{ij}\big|_S=[\partial_i\partial_j v_\phi]^{\TF}=\Dij v_\phi.
\end{equation}
In the TT sector, $\delta u_i=0$, $\delta g_{ij}\big|_{\TT}=a^2\chi_{ij}$, and $\delta\Gamma^0{}_{ij}\big|_{\TT}=a^2H\chi_{ij}+\frac{a^2}{2}\dot\chi_{ij}$, so
\begin{equation}
\delta\sigma_{ij}\big|_{\TT}=\left[a^2H\chi_{ij}+\frac{a^2}{2}\dot\chi_{ij}-Ha^2\chi_{ij}\right]^{\TF}=\frac{a^2}{2}\dot\chi_{ij}.
\end{equation}
Comparing with Eqs.~\eqref{eq:piScalarFinal} and \eqref{eq:piTTMain} establishes Eqs.~\eqref{eq:pi_sigma_scalar}--\eqref{eq:pi_sigma_viscous_final}.

\begin{acknowledgments}
The authors acknowledge funding from the Fundação para a Ciência e a Tecnologia (FCT) through the research grant UID/04434/2025.
FSNL also acknowledges support from the FCT Scientific Employment Stimulus contract with reference CEECINST/00032/2018.
\end{acknowledgments}

\bibliographystyle{apsrev4-2}
\bibliography{apssamp}

\end{document}